\documentclass[opendfm]{xlance/xlance}

\usepackage{amsthm}
\usepackage{tikz}
\usepackage{pifont}
\usepackage{booktabs}
\usepackage{algorithm}
\usepackage{algorithmic}
\usepackage{listings}
\usepackage{float}

\definecolor{xpfKeyword}{RGB}{35,92,160}
\definecolor{xpfField}{RGB}{125,76,160}
\definecolor{xpfVariable}{RGB}{17,128,117}
\definecolor{xpfComment}{RGB}{104,128,92}
\definecolor{xpfString}{RGB}{170,96,32}
\definecolor{xpfBg}{RGB}{248,250,252}
\definecolor{xpfFrame}{RGB}{218,226,235}

\lstdefinelanguage{xpf}{%
    alsoletter={_,-},
    morekeywords={protocol,version,entry,inputs,outputs,symbols,actors,policies,subflows,stages,type,role,source_policy,commit_policy,kind,profile,allowed_tools,reads,writes,uses,when,to,with,goto,call,return,outputs,passed,reason,python},
    morekeywords=[2]{actor,flow,handoff,judge,derive,derivation,string,str,array,bool,boolean,object,llm,human,tool,agent,working,output,auto,requires_human,true,false,null,exists},
    sensitive=true,
    comment=[l]{\#},
    morecomment=[l]{//},
    morestring=[b]",
    morestring=[b]',
    emph={research_question,final_brief,evidence_notes,draft_ready,implementation_plan,delivery_manifest,publish_ready,writer,reviewer,publisher,question,notes,draft,review_passed,review_reason,needs_evidence,collected_notes,topic,passed,reason,writer_only,reviewer_only,tool_or_human,compose_brief,concise_writer,drafting-base,intake,publish,revise},
    emphstyle=\color{xpfVariable},
}

\lstdefinestyle{plain}{%
    style=xlance@base,
    language=xpf,
    backgroundcolor=\color{xpfBg},
    rulecolor=\color{xpfFrame},
    basicstyle=\ttfamily\small\color{black!82},
    keywordstyle=\color{xpfKeyword}\bfseries,
    keywordstyle=[2]\color{xpfField}\bfseries,
    stringstyle=\color{xpfString},
    commentstyle=\color{xpfComment},
    numbers=none,
    xleftmargin=6pt,
    xrightmargin=6pt,
}

\lstdefinestyle{json}{%
    style=xlance@base,
    backgroundcolor=\color{xpfBg},
    rulecolor=\color{xpfFrame},
    basicstyle=\ttfamily\small\color{black!82},
    string=[s]{"}{"},
    stringstyle=\color{xpfVariable},
    comment=[l]{//},
    morecomment=[s]{/*}{*/},
    commentstyle=\color{xpfComment},
    literate=%
    *{:}{{\textcolor{black!55}{:}}}1
        {,}{{\textcolor{black!55}{,}}}1
        {\{}{{\textcolor{black!65}{\{}}}1
        {\}}{{\textcolor{black!65}{\}}}}1
        {[}{{\textcolor{black!65}{[}}}1
        {]}{{\textcolor{black!65}{]}}}1
        {true}{{\textcolor{xpfField}{true}}}4
        {false}{{\textcolor{xpfField}{false}}}5
        {null}{{\textcolor{xpfField}{null}}}4,
}

\newif\ifxflowappendixlistings
\pretocmd{\appendix}{\xflowappendixlistingstrue}{}{}
\BeforeBeginEnvironment{lstlisting}{\ifxflowappendixlistings\par\noindent\begin{minipage}{\linewidth}\fi}
\AfterEndEnvironment{lstlisting}{\ifxflowappendixlistings\end{minipage}\par\fi}

\usetikzlibrary{shapes.geometric, arrows.meta, positioning, fit, backgrounds, calc, decorations.pathreplacing, shadows}

\tikzset{
    box/.style={rectangle, rounded corners, draw=black!60, fill=blue!5,
                minimum width=2.5cm, minimum height=0.8cm, align=center,
                font=\small},
    stage/.style={rectangle, rounded corners, draw=orange!60, fill=orange!5,
                  minimum width=2cm, minimum height=0.7cm, align=center,
                  font=\small},
    actor/.style={ellipse, draw=green!60, fill=green!5,
                  minimum width=1.5cm, minimum height=0.7cm, align=center,
                  font=\small},
    arrow/.style={-{Stealth[length=2mm]}, thick, black!60},
    dataarrow/.style={-{Stealth[length=2mm]}, thick, dashed, blue!60},
    label/.style={font=\scriptsize\bfseries, align=center},
    container/.style={rectangle, rounded corners, draw=gray!40, fill=gray!5,
                      inner sep=0.5cm}
}

\theoremstyle{definition}

\definecolor{takeawayBg}{RGB}{250,252,255}
\definecolor{takeawayFrame}{RGB}{220,230,246}
\definecolor{takeawayAccent}{RGB}{76,119,190}
\definecolor{takeawayTitle}{RGB}{34,78,145}
\definecolor{takeawayTitleBg}{RGB}{235,243,255}

\newtcolorbox{takeaway}[1][]{
    enhanced,
    colback=takeawayBg,
    colframe=takeawayFrame,
    colbacktitle=takeawayTitleBg,
    coltitle=takeawayTitle,
    fonttitle=\bfseries,
    title={Key Takeaway},
    rounded corners,
    arc=3pt,
    boxrule=0.4pt,
    borderline west={2pt}{0pt}{takeawayAccent},
    left=9pt, right=9pt, top=6pt, bottom=6pt,
    #1
}

\crefname{assumption}{assumption}{assumptions}
\crefname{equation}{Eq.}{Eqs.}

\newcommand{\protocol}{\mathcal{P}}

\newcommand{\flow}{\mathcal{F}}
\newcommand{\handoff}{\mathcal{H}}

\title{\textit{XFlow}: An Executable Protocol Programming System for Reliable Multi-Agent Workflows}

\small{
\author[1,4]{\equalcontribution{Hanqi Li}}
\author[1]{\equalcontribution{Jing Peng}}
\author[1]{\equalcontribution{Zijian Wang}}
\author[1,2,3,4]{Lu Chen}
\author[1,3,4]{\correspondence{kai.yu@sjtu.edu.cn}{Kai Yu}}
}

\renewcommand\affiliation[2][]{\addtolist[#1]{#2}{\affiliationlist}{\affiliationformat}{\par}}
\small{
\affiliation[1]{X-LANCE Lab, School of Computer Science, Shanghai Jiao Tong University, Shanghai, China}
\affiliation[2]{Shanghai Innovation Institution, Shanghai, China}
\affiliation[3]{Jiangsu Key Lab of Language Computing, Suzhou, China}
\affiliation[4]{Suzhou Laboratory, Suzhou, China}
}
\abstract{
LLM-based multi-agent systems increasingly coordinate planning, reasoning, tool use, and human interaction, yet their reliability remains limited.
A central source of this limitation is the underspecified prompt--harness boundary.
Current systems lack a principled way to decide which workflow commitments should remain in prompts and which should become harness structure.
We present \textbf{XFlow}, an executable protocol programming system for reliable multi-agent workflows, and \textbf{XPF} (XFlow Protocol Format), its domain-specific protocol programming language.
XFlow occupies a middle position between prompt-only orchestration and markup-like workflow descriptions.
XPF remains readable as a literate protocol, but it is compiled and executed as a program.
Its design keeps informal semantic work inside actors while moving selected commitments into harness structure that can be checked, preserved, and enforced.
At runtime, XFlow stages uncertainty through lifecycle-governed symbols, which are typed state cells with validation and commit states.
Actor outputs are mediated before they become shared state, instead of spreading through prompts, transcripts, or implicit memory.
Our experiments cover Constrained Interaction, Long-Context Reasoning, and Agentic Software Engineering.
They show that XFlow improves reliability by making constraints, evidence handling, and process requirements explicit and enforceable.
}

\newcommand{\titlepageteaser}{%
    \vspace{0.35cm}
    \begin{center}
        \includegraphics[width=0.96\linewidth]{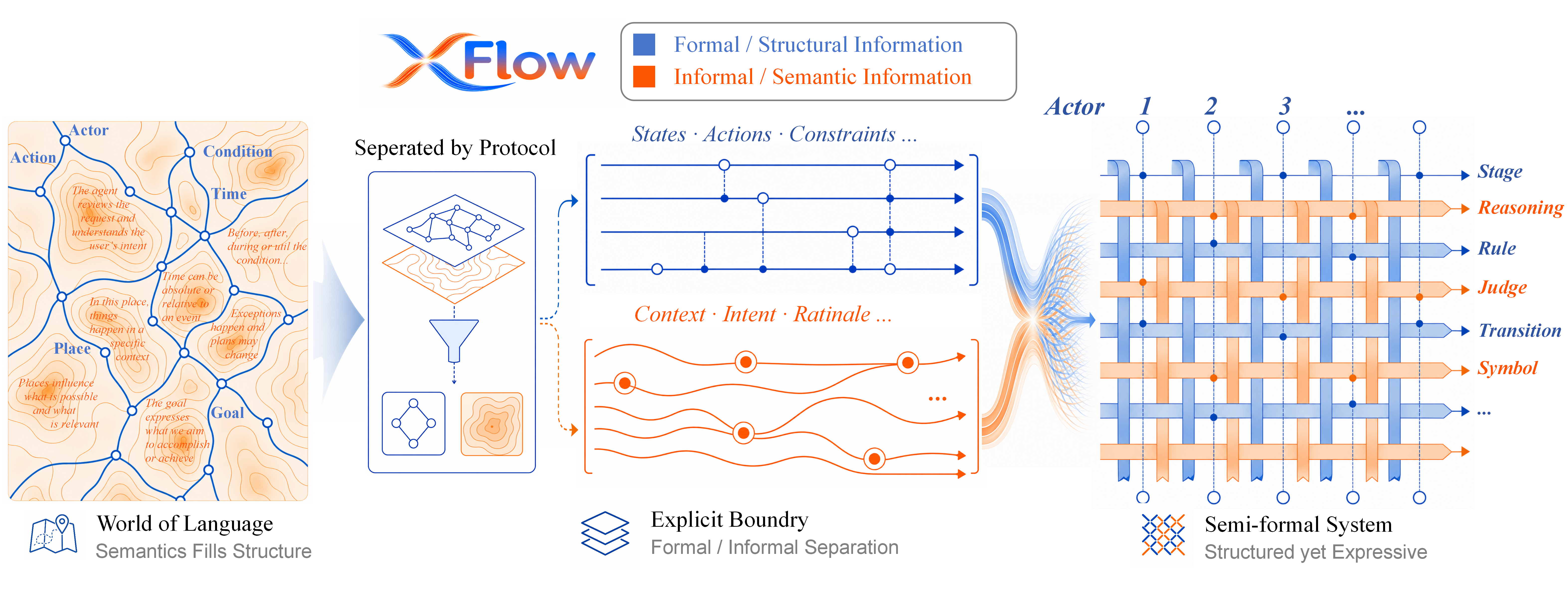}
    \end{center}
    \vspace{0.2cm}
}

\begin{document}

\publishdate{\today}
\maketitle
\titlepageteaser

\tableofcontents
\newpage

\section{Introduction}
\label{sec:introduction}

Large language models (LLMs) are increasingly used as autonomous agents that plan, reason, and act on behalf of users~\citep{wei2022chain, yao2023react, xi2025rise}.
This flexibility has encouraged a shift from single-agent prompting to \emph{multi-agent systems}, where specialized LLM actors work together through structured workflows~\citep{park2023generative, hong2023metagpt}.
The shift expands what agents can do, but it also exposes a deeper orchestration problem.
A hallucinated fact, malformed output, or misinterpreted instruction may be tolerable when isolated inside one agent.
In a multi-agent pipeline, the same error can become shared state, trigger downstream actions, and corrupt later decisions~\citep{ji2023survey}.
Reliability therefore depends on more than reducing model errors.
It also depends on how a workflow passes information between actors and keeps key knowledge and constraints stable.

This is a problem of boundary.
In agent applications, it appears operationally as the division between the \emph{prompt} and the \emph{harness}.
More generally, it is the formal/informal boundary between governable commitments and informal semantic work.
Governable commitments include constraints, state interfaces, policies, and process obligations.
This boundary is hard to define precisely because knowledge, logic, rules, and judgment often mix in practice.
Yet without mechanisms for placing and governing it, developers have little basis for improving it.
Today, many commitments that could be symbolized and structured remain embedded in natural-language prompts, skills, or instructions.
Actors must then repeatedly interpret and enforce knowledge or constraints that the surrounding system could handle directly.
As a result, governable knowledge and constraints become things an actor must read, remember, and follow.
They do not become explicit protocol state or rules that a harness can inspect, preserve, and enforce.
Agent developers need explicit control over this placement: what stays inside the prompt, and what becomes part of the harness.

The Chomsky hierarchy~\citep{chomsky1956three} helps explain why this boundary remains obscure.
It frames languages by their degree of structure and mechanical control.
Current orchestration frameworks tend to sit near two ends of this spectrum.
Markup- or configuration-based frameworks provide external structure, but they mainly describe workflow shape.
They say less about where knowledge and constraints leave the prompt and become harness responsibilities.
Prompt-based frameworks sit near the natural-language end.
They preserve expressive flexibility, but leave potentially formalizable knowledge and constraints inside instructions.
Because the middle ground is weakly developed, systems lack a language for drawing, testing, and adjusting the prompt--harness boundary.

We present \textbf{XFlow}, an executable protocol programming system for reliable multi-agent workflows, and \textbf{XPF} (XFlow Protocol Format), its domain-specific protocol programming language.
XFlow occupies a middle ground between prompt-only orchestration and markup-like workflow descriptions.
XPF remains readable as a literate protocol, but it is compiled and executed as a program.
It is semi-formal in this sense: readable to authors, but structured enough for the harness to compile and enforce.
This design makes the prompt--harness boundary explicit.
Protocol authors can keep informal semantic work inside actors while externalizing selected commitments into harness structure that can be checked, preserved, and enforced.
At runtime, XFlow stages uncertainty through lifecycle-governed symbols, which are typed state cells with validation and commit states.
Actor outputs must pass through this mediation before they become shared state, instead of spreading through prompts, transcripts, or implicit memory.
XFlow therefore does not replace prompt work with harness logic. It gives authors explicit control over which commitments become governable before actor outputs influence downstream decisions.

Empirically, we evaluate XFlow on three benchmark classes: Constrained Interaction, Long-Context Reasoning, and Agentic Software Engineering.
Across these settings, XFlow improves performance and reliability while preserving actor flexibility.
Together, the results show that XFlow provides more than benchmark-specific gains.
It gives developers an operational mechanism for drawing and governing the prompt--harness boundary.
XFlow lifts commitments into explicit executable protocol structures.
This makes them inspectable, preservable, and enforceable across heterogeneous multi-agent workflows, while leaving informal semantic work to the actors.

This paper makes three contributions:
\begin{itemize}
  \item We introduce \textbf{XFlow}, an executable protocol programming system for reliable multi-agent workflows. XFlow treats the prompt--harness boundary as an explicit design object, so protocol authors can decide which commitments remain with actors and which become enforceable workflow structure.
  \item We design \textbf{XPF}, a semi-formal protocol format, together with compiler and runtime support. XPF turns symbols, actor interfaces, flow rules, handoffs, and lifecycle policies into executable protocol structures that can be checked, preserved, and enforced.
  \item We evaluate XFlow on Constrained Interaction, Long-Context Reasoning, and Agentic Software Engineering benchmarks. The results show how protocol externalization improves reliability while preserving actor flexibility.
\end{itemize}

The rest of this paper is organized as follows.
Section~\ref{sec:philosophy} presents the design philosophy.
Section~\ref{sec:method} presents the XFlow architecture and design.
Section~\ref{sec:experiments} reports experimental results.
Section~\ref{sec:cloud_edge} discusses cloud--edge coordination in real-world agent workflows.
Section~\ref{sec:related_work} surveys related work.
Section~\ref{sec:conclusion} concludes.

\section{Design Philosophy}
\label{sec:philosophy}

\begin{figure}[t]
\centering
\includegraphics[page=1,width=\textwidth]{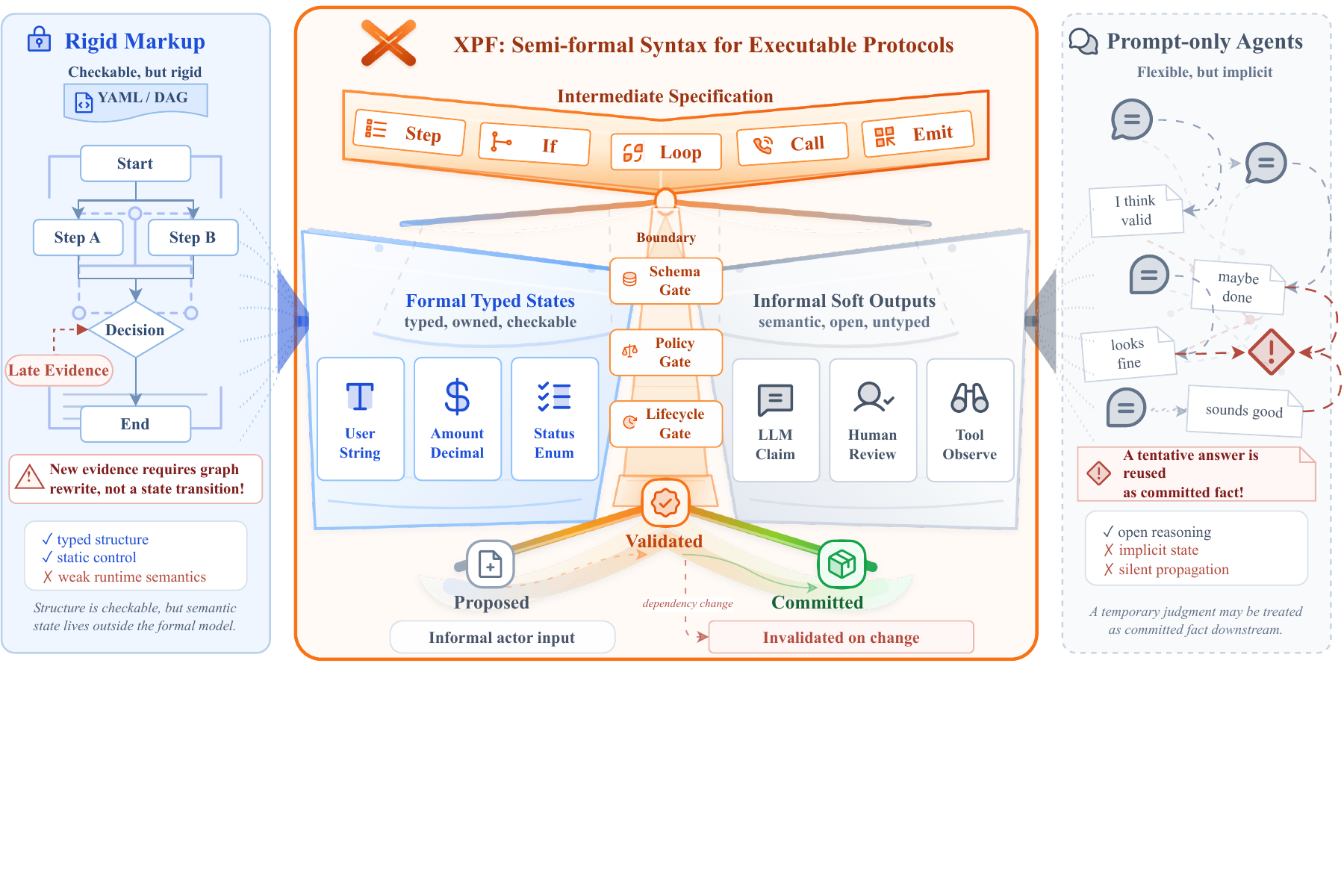}
\caption{Design philosophy of XFlow. Rigid formal orchestration is checkable but weak at representing semantic state, while informal prompt-only orchestration is flexible but leaves state and commitments implicit. XPF occupies a semi-formal protocol-programming regime: typed symbols, schemas, policies, and lifecycle gates mediate actor outputs before they become shared workflow state.}
\label{fig:philosophy}
\end{figure}

Figure~\ref{fig:philosophy} summarizes the design logic developed in this section.
XFlow starts from a practical observation: workflow commitments that now sit implicitly in prompts can move into harness structure, while informal semantic work should remain with actors.
This observation turns the prompt--harness boundary into a design object rather than an implicit convention.
The section first positions XPF as a semi-formal middle ground between markup-like workflows and prompt-only orchestration.
It then explains how XPF controls boundary placement and how typed symbols govern values that cross the boundary.

\subsection{Occupying an Intermediate Regime}
\label{sec:phil_chomsky}

Current LLM-agent systems often rely on two orchestration extremes.
We use the Chomsky hierarchy~\citep{chomsky1956three} as a lens for the syntactic complexity of workflow specifications, not as a literal classification of whole executions.
At one end, markup- or configuration-style orchestrators express workflow control through structured objects such as chains, graphs, nodes, tool interfaces, and configuration fields.
This makes parts of the harness parseable and executable, but the grammar usually governs control structure rather than the placement of semantic commitments between prompts and harness state.
At the other end, prompt-centered orchestrations express rich task knowledge, policies, exceptions, and coordination rules in natural language.
This preserves semantic flexibility, but leaves those commitments difficult for the harness to inspect, preserve, or enforce as protocol state.
Both styles may combine prompts and harness logic.
What remains weakly specified is the boundary that decides which commitments stay as actor-interpreted instructions and which become declared protocol state or rules.

XFlow is designed to occupy this missing middle ground.
Its protocol programming language, XPF, is \emph{literate and executable}: readable as human-facing documentation, yet structured enough for the harness to parse and execute.
At this semi-formal level, the purpose of structure is modest but important.
It gives the workflow a stable protocol surface, so authors can separate what actors should interpret from what the runtime should carry.
This intermediate position prepares the ground for the boundary question addressed next.

\subsection{Making the Formal/Informal Boundary Explicit}
\label{sec:phil_boundary}

The intermediate position of XPF is useful because it explicitly constructs and exposes the formal/informal boundary.
In practical agent systems, this is the prompt--harness boundary.
The question is simple: which parts of a workflow should leave in the prompt, and which parts should the harness name, check, preserve, and enforce?
XPF makes this question part of the protocol itself rather than leaving it as an implicit convention.

It does so through its own structure.
Prose remains readable task context for actors.
Semantic blocks declare the executable parts of the workflow, including actor interfaces, flow rules, and handoffs.
Symbols name the state that may cross from actor work into shared workflow state.
In this way, XPF exposes the boundary while authors write the protocol: informal semantic work stays in actor-facing text, while governable commitments become declared protocol structure.

This explicit boundary decides where responsibility lives.
Actors can still interpret, reason, write, judge, and use tools.
The harness does not replace that work.
It controls when actor outputs may affect shared state.
An actor turn can do so only through declared outputs, and those outputs must pass schema, provenance, and commit checks through the symbol lifecycle.

\subsection{Governing Uncertainty with Symbols}
\label{sec:phil_uncertainty}

For XFlow, a reliable multi-agent system does not eliminate uncertainty.
Instead, it locates, constrains, and governs uncertainty.
Uncertainty enters through prompt-guided actors, and it becomes risky when actor outputs cross into harness-governed state.
The formal side of the boundary cannot make these outputs certain, but it can control how they become shared and reusable.
The prompt--harness boundary is therefore also an uncertainty boundary.
It collects tentative actor interpretations into explicit objects rather than letting them spread through prompts, transcripts, or implicit memory.
As in transactional semantics in databases~\citep{gray1981transaction}, an uncommitted write should not be treated as committed data, and partial or conflicting updates should not corrupt consistent state.

The object that realizes this boundary is the \emph{symbol}.
A symbol is not just a variable name in a prompt.
It is a typed runtime state cell through which the harness can hold externalized knowledge, constraints, and actor outputs.
Each symbol carries a schema, source policy, commit policy, provenance, dependencies, and lifecycle state.
Actor outputs first enter the system as \textsc{Proposed} symbol values.
They become \textsc{Validated} only after schema and contract checks.
They become \textsc{Committed} only after the relevant commit policy is satisfied.
When an upstream symbol changes, dependent symbols can be invalidated or recomputed by the reactive runtime.
Symbols therefore absorb uncertainty and reusable structure into protocol state before they can influence downstream work.

\begin{takeaway}
XFlow's design rests on three principles.
First, agent orchestration needs an intermediate protocol programming regime between markup-like workflows and prompt-only control.
Second, this regime should make the prompt--harness boundary explicit, so commitments that can be formalized move from prompts into the harness while informal semantic work stays with actors.
Third, reliability means managing uncertainty by absorbing actor outputs into typed symbols whose lifecycle determines when they may influence downstream state.
\end{takeaway}

\section{XFlow Architecture}
\label{sec:method}
\begin{figure}[t]
\centering
\includegraphics[width=\textwidth]{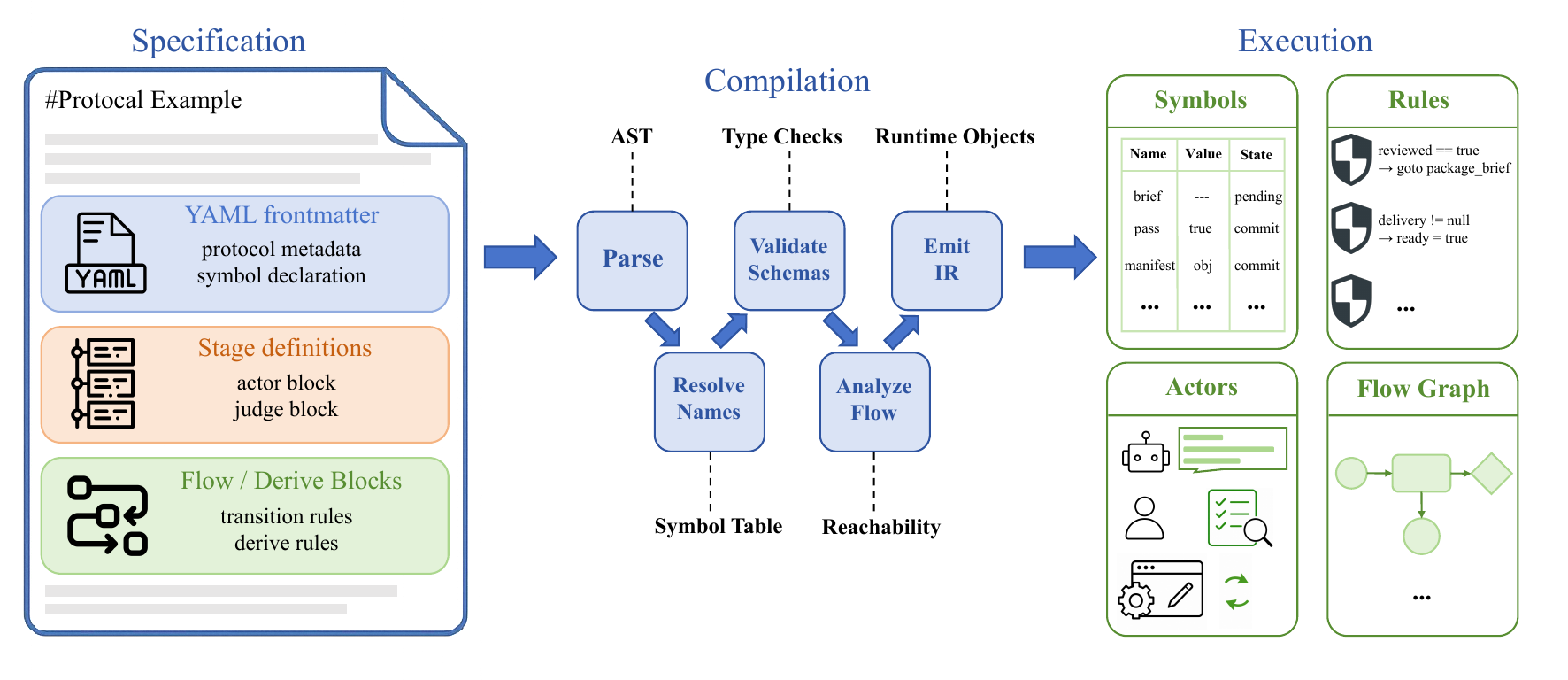}
\caption{XFlow Architecture. A single literate artifact exposes human-readable protocol text. The compiler recovers symbols, stages, flow graphs, a typed intermediate representation (IR), and runtime objects for execution.}
\label{fig:xpf-example}
\end{figure}
Building on the design philosophy above, this section defines XFlow as an executable protocol language system for multi-agent workflows.
Its architecture has three layers.
The outer Specification layer uses XPF, a human-facing protocol language.
Its syntax follows a \textit{``What You Say Is What You Execute''} discipline: the commitments authors write are the commitments the system later compiles and runs.
The middle Compilation layer parses XPF and resolves symbols, stages, actors, flow rules, and handoffs.
It then lowers them into a typed intermediate representation (IR), checked agent code, and runtime objects.
The Execution layer uses the XFlow runtime to manage actor calls, symbol lifecycles, tool gates, recovery frames, and auditable state transitions.
As shown in Figure~\ref{fig:xpf-example}, these layers connect literate protocol text to executable agent behavior without separating documentation from orchestration.

\subsection{Specification Phase}

\begin{table}[t]
\centering
\small
\setlength{\tabcolsep}{4.5pt}
\renewcommand{\arraystretch}{1.16}
\begin{tabularx}{\textwidth}{>{\centering\arraybackslash}p{0.12\textwidth} >{\raggedright\arraybackslash}p{0.18\textwidth} X >{\raggedright\arraybackslash}p{0.25\textwidth}}
\toprule
\rowcolor{xpfFrame}
\textbf{Notation} & \textbf{Name} & \textbf{Definition} & \textbf{Example} \\
\midrule
$\protocol$ & Protocol & The complete workflow contract. & An expense-reimbursement workflow from receipt submission to payment. \\
$\Sigma$ & Symbol space & Shared workflow state and lifecycle constraints. & Employee name, receipt images, total amount. \\
$\Delta$ & Derivations & Computed values and their dependencies. & Total amount is computed from the submitted receipt. \\
$\Lambda$ & Actor interfaces & Actor-stage bindings with visible inputs and writable outputs. & An employee uploads receipts. \\
$\flow$ & Control flow & Allowed stage transitions and protocol calls. & Missing receipts return to the employee. \\
$\handoff$ & Responsibility transfer & Explicit changes in who continues the work. & After submission, responsibility moves to the manager. \\
\bottomrule
\end{tabularx}
\caption{XFlow protocol abstraction and its main components.}
\label{tab:protocol-abstraction}
\end{table}

In the Specification phase, authors write the protocol document that XFlow will later compile and execute.
This document is intentionally readable like a workflow description.
XPF (XFlow Protocol Format) lets prose state task intent.
YAML frontmatter and semantic code blocks name the commitments that must become executable.
We abstract the resulting protocol as $\protocol = (\Sigma, \Delta, \Lambda, \flow, \handoff)$.
Table~\ref{tab:protocol-abstraction} summarizes these components.
Together, they define the protocol structure: $\Sigma$ defines shared state, $\Delta$ links values through derivations, $\Lambda$ binds actors to readable and writable channels, $\flow$ orders stages and protocol calls, and $\handoff$ records responsibility transfer.
Through these components, the protocol defines an agent system.
It assigns informal semantic work to actors and declares the governed channels through which actor outputs may change shared state, trigger later stages, or transfer responsibility.
The following subsections describe how each component is written in XPF and how it helps build the executable protocol.
Although the components are simple in isolation, their references resolve against the surrounding protocol context.
A symbol name, actor alias, predicate, or transition target has meaning only relative to nearby declarations.
This gives XPF a context-sensitive character.
It is neither unconstrained natural language nor a closed configuration schema, but a compositional protocol language whose small parts can define complex agent behavior.
Appendix~\ref{app:xpf-syntax} gives the detailed XPF syntax reference.

\subsubsection{Symbol Space ($\Sigma$)}

The symbol space gives actors a shared state layer without treating arbitrary transcript text as memory.
It identifies the values an actor may receive, produce, or rely on, and lets the runtime distinguish tentative outputs from committed workflow facts.
In XPF, $\Sigma$ is the protocol-level vocabulary that all other components reference.
Symbols can be read by actors, written by actors, computed by derivations, tested by flow predicates, and scoped through child protocols.
As a result, a small symbol declaration can participate in many later constraints.

To define $\Sigma$, authors write a \texttt{symbols} map in the protocol header.
Each symbol declares a type.
It may also declare its role, who may propose it, and when it may become committed state.
Later actor, derive, flow, and handoff blocks use these names as their state references.
Changing a symbol's type or lifecycle policy therefore changes how multiple parts of the protocol compose around it.

\par\noindent\begin{minipage}{\linewidth}
\begin{lstlisting}[style=plain]
symbols:
  question:              # stable task input
    type: string
    role: input
    commit_policy: immutable
  evidence_notes:        # reused by actors, rules, and flow
    type: array[string]
    role: working
    source_policy: tool_or_human
    commit_policy: auto
  draft:                 # writer output channel
    type: string
    role: working
    source_policy: writer_only
    commit_policy: auto
  review_passed:         # decision used by flow and handoff
    type: bool
    role: judge_output
    source_policy: reviewer_only
    commit_policy: requires_human
\end{lstlisting}
\end{minipage}

\subsubsection{Derivations ($\Delta$)}

Derivations move mechanical bookkeeping out of actor prompts and into deterministic protocol logic.
They let the system compute facts such as readiness, freshness, coverage, and routing conditions instead of asking an actor to remember them.
In XPF, $\Delta$ is the dependency layer that connects symbols and supports invalidation or recomputation when upstream state changes.
This layer is more general than value forwarding.
Derived symbols may be computed by built-in predicates, inline expressions, or external package code.
The protocol can therefore encode simple guards or larger deterministic checks without moving them into prompts.

To define $\Delta$, authors mark derived symbols in the header and add \texttt{derive} blocks where the computation becomes relevant.
A derive block names its output with \texttt{writes}, lists dependencies with \texttt{from}, and supplies either a small expression or a package-local function.
Predicates such as \texttt{exists} are provided by the XPF runtime predicate library, while larger rules can be referenced through a \texttt{python} field such as \texttt{rules.readiness:draft\_ready}.
Unlike actor blocks, these declarations describe deterministic protocol logic rather than informal semantic work.

\par\noindent\begin{minipage}{\linewidth}
\begin{lstlisting}[style=plain]
symbols:
  draft_ready:           # computed guard for later flow
    type: bool
    role: derived
    from: [draft]        # dependency tracked by runtime

```derive
writes: draft_ready      # derived symbol to update
from: [draft]            # recompute if draft changes
expr: exists(draft)      # built-in runtime predicate
```

```derive
writes: draft_ready      # same output via package code
from: [draft]
python: rules.readiness:draft_ready
```
\end{lstlisting}
\end{minipage}

\subsubsection{Actor Interfaces ($\Lambda$)}

Actor interfaces determine what each LLM or human actor is allowed to see and change.
They keep reasoning, evidence selection, and wording open, but give each actor turn explicit inputs, outputs, and authority.
In XPF, $\Lambda$ is the binding layer that connects named actors to stages through readable and writable symbol channels.
An actor name may point to an LLM profile, a human interface, or another external package adapter.
The protocol therefore composes open-ended behavior with a fixed invocation boundary instead of embedding implementation details in the workflow text.

To define $\Lambda$, authors declare reusable actors and bind them to stages with \texttt{actor} or \texttt{judge} blocks.
The block names the actor, lists readable symbols, and lists writable symbols or mapped judgment outputs.
The surrounding prose remains actor context.
The block defines what the runtime may expose to the actor and accept in return.
The same actor can then be reused in many stages with different read--write authority.

\par\noindent\begin{minipage}{\linewidth}
\begin{lstlisting}[style=plain]
actors:
  writer:                # reusable LLM actor
    kind: llm
    profile: concise_writer # external invocation profile

## Draft Brief {#draft-brief}
Write a brief answer to the question.

```actor
uses: writer             # actor implementation to invoke
reads: [question]        # visible input symbols
writes: [draft]          # only allowed state output
```
\end{lstlisting}
\end{minipage}

\subsubsection{Control Flow ($\flow$)}

Control flow determines how work continues after an actor turn.
It prevents the next step from being inferred from free-form text by declaring which stages or child protocols the runtime may enter.
In XPF, $\flow$ is the transition layer that connects state predicates to legal continuations.
Guards may combine derived symbols, actor judgments, child-protocol returns, and built-in predicates.
Even compact flow blocks can therefore express loops, retries, escalation paths, and hierarchical workflows.

To define $\flow$, authors write \texttt{flow} blocks whose \texttt{goto} and \texttt{call} statements are guarded by predicates over declared symbols.
Targets must resolve to stages or child protocols, and call returns must map child outputs back into parent symbols.
Thus the protocol controls both ordinary stage movement and nested workflow calls while keeping the allowed continuation space explicit.

\par\noindent\begin{minipage}{\linewidth}
\begin{lstlisting}[style=plain]
```flow
goto review-brief when draft_ready == true       # proceed when ready
goto draft-brief when draft_ready != true        # retry drafting path
call evidence/collect when exists(evidence_notes) != true # child protocol
  return:
    collected_notes: evidence_notes              # child output mapping
```
\end{lstlisting}
\end{minipage}

\subsubsection{Responsibility Transfer ($\handoff$)}

Responsibility transfer determines who owns the next action after a condition is met.
It differs from control flow: a transition says where execution may go, while a handoff says which actor or role should continue the work.
In XPF, $\handoff$ is the responsibility layer that makes approval, review, escalation, and human intervention visible as protocol state.
Handoffs use the same symbol and predicate context as flow rules.
Responsibility can therefore depend on actor outputs, derived checks, or external review results without being hidden inside natural-language instructions.

To define $\handoff$, authors attach \texttt{handoff} blocks to the relevant stage.
Each block names a guard condition, the receiving actor or role, and optionally a reason that remains visible to readers, operators, and the compiler.
At runtime, the handoff updates responsibility only when the declared condition is satisfied, and the resulting ownership change becomes part of auditable protocol state.

\par\noindent\begin{minipage}{\linewidth}
\begin{lstlisting}[style=plain]
```handoff
when: review_passed == true              # transfer only after approval
to: publisher                            # actor or role that continues
reason: approved brief is ready for release
```
\end{lstlisting}
\end{minipage}

\subsection{Compilation Phase}

The Compilation phase is the middle layer of XFlow's executable protocol language system.
Its task is to turn the XPF protocol document into executable agent code and runtime objects before any actor or tool runs.
Unlike Specification, this phase is not mainly an authoring interface.
It is a sequence of mechanical passes that make the protocol executable.
Following Figure~\ref{fig:xpf-example}, the compiler first parses the literate surface, resolves names and paths, and checks interfaces and policies.
It then lowers the checked structures into typed objects and emits a typed IR package.
The output is a checked execution contract.
It preserves the author's protocol while removing ambiguity about what the runtime may execute.
Appendix~\ref{app:ir-schema} provides the detailed IR schema, lowering steps, and static check summaries.

\subsubsection{Parse the Literate Surface}

Compilation begins by separating the mixed XPF document into parts that later passes can reason about.
YAML frontmatter becomes protocol-level declarations, Markdown headings become stage scopes, prose remains documentation and actor context, and fenced semantic blocks become executable declarations attached to the nearest stage.
This pass changes a readable protocol document into a parsed surface: still close to the source text, but no longer just Markdown.

\par\noindent\begin{minipage}{\linewidth}
\begin{lstlisting}[style=plain]
# source surface
symbols:
  draft: {type: string}

## Draft Brief {#draft-brief}
Write the first answer.

```actor
uses: writer
reads: [question]
writes: [draft]
```

# parsed surface
{
  "symbols": {"draft": {"type": "string"}},
  "stages": [{
    "id": "draft-brief",
    "prose": "Write the first answer.",
    "blocks": [{"type": "actor", "uses": "writer",
                "reads": ["question"], "writes": ["draft"]}]
  }]
}
\end{lstlisting}
\end{minipage}

\subsubsection{Resolve Names and Check Commitments}

The next passes give every reference a stable meaning and reject inconsistent protocols.
The compiler checks symbol names against the symbol table, stage targets against the recovered stage map, and actor names against actor and profile specifications.
It resolves tool, schema, derivation, and child-protocol references relative to the package.
It then checks that actor reads and writes are legal.
It checks that judge outputs bind to declared symbols, derivations depend on known symbols, and flow targets exist.
It also checks that call returns match child-protocol outputs and that source or commit policies are not violated.

This is the static side of the prompt--harness boundary.
The compiler does not predict what an LLM or human will say; it verifies the channels through which that content may affect shared state.
The compiler catches errors such as missing symbols, unresolved stages, invalid Python rule references, incompatible schemas, and bad call bindings before execution.

\par\noindent\begin{minipage}{\linewidth}
\begin{lstlisting}[style=plain]
# unresolved flow text
goto review-brief when draft_ready == true
call evidence/collect when exists(evidence_notes) != true
  return:
    collected_notes: evidence_notes

# resolved and checked references
{
  "goto": {
    "target_stage_id": "brief/review-brief",
    "predicate_symbols": ["draft_ready"]
  },
  "call": {
    "target_protocol": "brief/evidence/collect",
    "uses_builtin_predicate": "exists",
    "return_bindings": {"collected_notes": "evidence_notes"}
  }
}
\end{lstlisting}
\end{minipage}

\subsubsection{Lower to Typed IR}

After parsing, resolution, and checking, the compiler lowers the protocol into the typed objects used during execution.
A symbol declaration becomes an \texttt{IRSymbol}.
A heading and its blocks become an \texttt{IRStage}.
An actor block becomes an actor interface with read and write sets.
A derivation becomes a dependency edge.
A flow rule becomes a checked transition or scoped protocol call.
The resulting IR package contains the stage graph, symbol table, actor interfaces, policies, dependency graph, subflow imports, and runtime metadata.

This lowering step changes the kind of object the system handles.
The runtime no longer needs to reparse XPF or infer the next step from prompt text.
It receives a typed execution artifact whose symbols, actor boundaries, derived dependencies, transitions, calls, and handoffs are already explicit.

\par\noindent\begin{minipage}{\linewidth}
\begin{lstlisting}[style=plain]
{
  "protocol_id": "brief",
  "entry_stage_id": "brief/draft-brief",
  "symbols": {
    "draft": {"type": "string", "source_policy": "writer_only"},
    "draft_ready": {"type": "bool", "role": "derived"}
  },
  "stages": {
    "brief/draft-brief": {
      "agent": {"ref": "writer", "reads": ["question"], "writes": ["draft"]},
      "derives": [{"writes": "draft_ready", "from": ["draft"]}],
      "flow_rules": [{"goto": "brief/review-brief", "when": "draft_ready == true"}]
    }
  }
}
\end{lstlisting}
\end{minipage}

\subsection{Execution Phase}

The Execution phase is the third layer of XFlow's executable protocol language system.
The runtime machinery implements this phase.
Here, execution names the phase in the architecture, while the runtime names the mechanism that drives the compiled agent code and enforces the compiler's contract.
Actor calls follow declared interfaces.
Symbol writes pass through lifecycle checks.
Protocol state mediates tool and human gates, while recovery frames preserve resumable context.
Unlike Specification and Compilation, this phase must coexist with uncertainty because LLM and human actors may produce useful, incomplete, surprising, or delayed outputs.
XFlow therefore does not treat runtime execution as a free-form transcript.
It stages actor outputs before commitment and records enough context to resume or audit long-running work.
The runtime governs three crossing points where informal behavior can affect the formal workflow: actor invocation, symbol state updates, and session recovery.
These crossing points match the specification-level boundaries.
Interfaces mediate actor action, symbol lifecycles mediate state changes, and scoped session records mediate pauses or nested calls.
Appendix~\ref{app:actor-output} details the actor output envelope, Appendix~\ref{app:session-format} details symbol state and lifecycle records, and Appendix~\ref{app:recovery-reference} describes session recovery records.

\subsubsection{Interface-Mediated Actor Action}

\begin{figure}[!t]
\centering
\includegraphics[width=\linewidth]{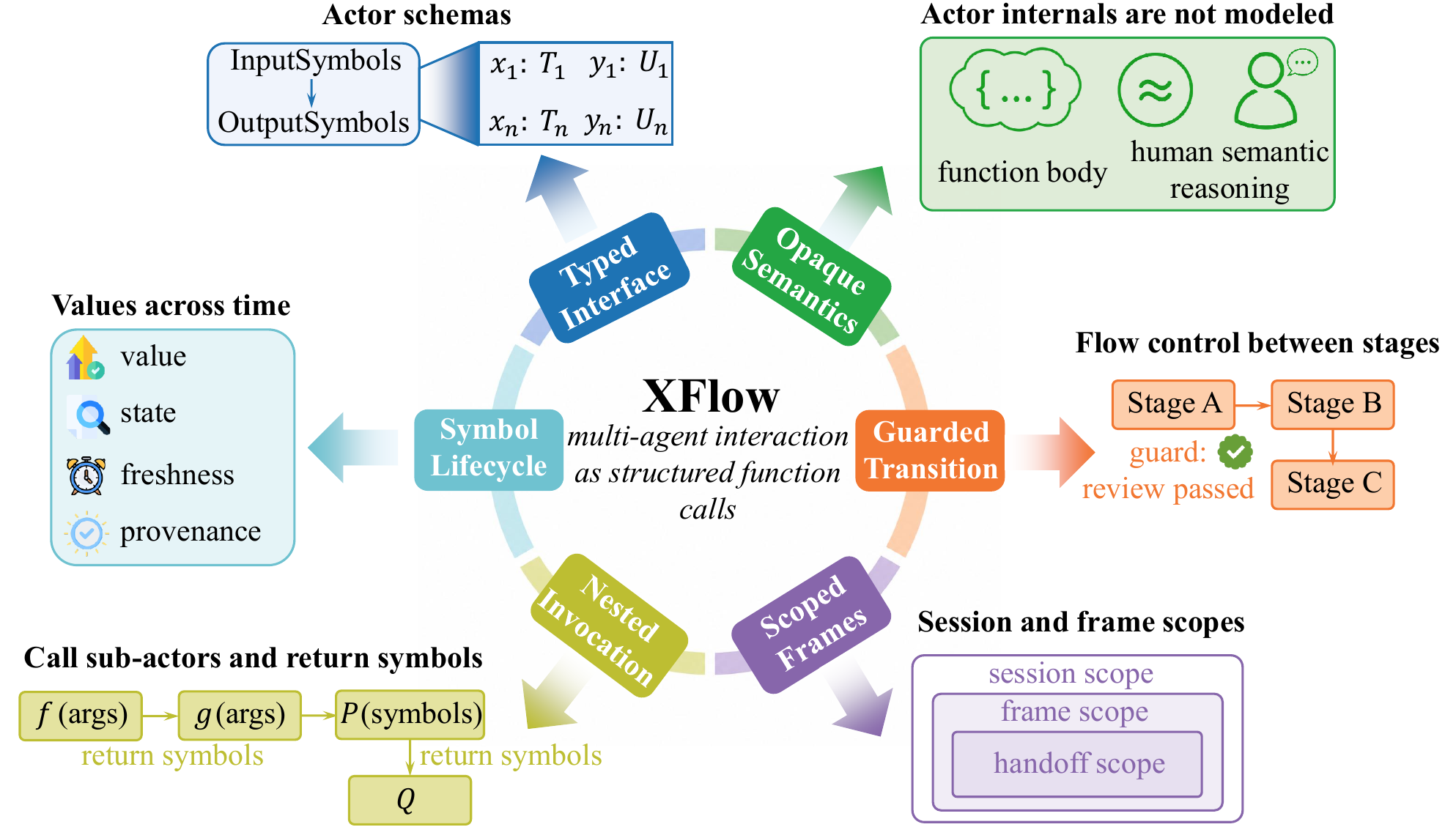}
\caption{Actors as typed function calls in XFlow. Actor reasoning remains unconstrained, but each invocation crosses a typed boundary with declared inputs, outputs, and terminal events. Flow rules connect these calls into workflows, and child protocols behave as scoped calls.}
\label{fig:function-view}
\end{figure}

Actors form the informal side of the system.
An LLM actor may synthesize text, call tools, or revise an answer, while a human actor may provide judgment, approval, or external input.
The runtime does not formalize this internal reasoning.
Instead, it wraps each invocation in a typed boundary.
Prompts are assembled from declared inputs and profile context.
Responses are parsed into declared outputs.
Terminal events such as returns, calls, or handoffs are interpreted as runtime events rather than free-form transcript text.

At a high level, XFlow treats an actor as a function-like unit of execution: the actor body remains informal, but the invocation boundary is typed and governed (Figure~\ref{fig:function-view}). 
This boundary gives LLM and human actors the same operational status.
Both are invoked through a stage interface, both receive only the symbols made visible by the protocol, and both must return values through declared output channels.
The difference lies in how the actor body is implemented.
An LLM actor may use prompt assembly, schema-guided decoding, retries, and tools, while a human actor may use interactive input, asynchronous response, or approval screens.
The runtime treats these differences as actor implementations rather than as changes to the protocol semantics.

\subsubsection{Governed State Commitment}

The runtime stores workflow values as lifecycle-governed symbol cells.
An actor output may create a proposed value, but it becomes reliable state only after schema validation, provenance recording, and any required commit policy check.
When a committed value changes, dependency tracking marks derived values stale and schedules recomputation, following a reactive execution model inspired by functional reactive programming~\citep{elliott1997functional, bainomugisha2013survey}.
Transactions keep these updates atomic.
If validation or recomputation fails, the symbol store returns to its previous consistent state.

The lifecycle is what turns uncertainty into an explicit runtime object.
A candidate output can be stored, inspected, rejected, or revised without being confused with committed state.
Source policies determine whether the producer was allowed to propose the value.
Schema checks determine whether the value has the declared shape.
Commit policies determine whether additional approval is required before downstream stages may rely on it.
Reactive invalidation then prevents stale derived values from silently surviving after their dependencies change.

\subsubsection{Persisting Execution Context}

Runtime execution may pause, branch, or enter a child protocol.
XFlow records these situations as scoped frames containing visible symbols, pending outputs, event-log positions, and blocked human gates.
Session state combines these frames with the symbol store and event log.
Execution can then resume without reconstructing context from an informal transcript.
Operators can also audit who proposed values, which checks ran, and why execution is blocked.

Protocol calls use the same frame mechanism: a child protocol receives only mapped inputs and returns only declared outputs.
This prevents accidental parent-state mutation by name coincidence and keeps nested workflows recoverable.

\begin{takeaway}
XFlow has a three-phase architecture.
Authors specify a workflow as an XPF protocol.
The compiler turns that protocol into checked executable runtime objects.
The runtime executes those objects while enforcing actor, state, and recovery boundaries.
The protocol is therefore not separate documentation.
It is the source that gets compiled and run.
\end{takeaway}

\section{Experimental Results}
\label{sec:experiments}

We evaluate XFlow on three benchmark settings that stress different parts of the protocol-to-execution architecture.
$\tau^3$-bench~\citep{yao2024taubench,barres2025tau2} tests Constrained Interaction.
CorpusQA~\citep{lu2026corpusqa} tests Long-Context Reasoning.
SWE-bench Verified~\citep{jimenez2024swebench,openai2024swebenchverified} tests Agentic Software Engineering.
Each setting places different requirements on the XFlow protocol.
Unless otherwise noted, experiments use Qwen3.5-9B~\citep{qwen2026qwen35} and DeepSeek-V4-Flash~\citep{deepseekai2026deepseekv4} as backbone models.

\subsection{Constrained Interaction: \texorpdfstring{$\tau^3$-bench}{tau3-bench}}
\label{sec:tau3-analysis}

The $\tau^3$-bench evaluation studies the relationship between agent correctness and reliability under explicit domain constraints.
A trajectory may complete the user-facing task while still violating a policy constraint, such as an eligibility rule, refund condition, required disclosure, or domain-specific operating procedure.
We therefore report \emph{Compliant} as a reliability metric derived from each task's constraint document.
We extract constraints from the benchmark-provided policy text and check them against the agent trajectory, in addition to the benchmark's native task-success criterion.
Following the $\tau$-bench convention, pass$^{1}$ denotes the benchmark's single-trial task-success metric, i.e., the $k=1$ case of pass$^{k}$.
We report \emph{Conditional pass} as the joint rate at which a trajectory both satisfies pass$^{1}$ and remains compliant.

Table~\ref{tab:tau3-results} shows that XFlow improves compliance while preserving comparable task success under the relevant constraints.
With Qwen3.5-9B, XFlow raises compliance from 96.5\% to 100.0\% in Retail, from 91.8\% to 100.0\% in Airline, and from 77.2\% to 82.5\% in Telecom.
pass$^{1}$ remains close: it decreases slightly in Retail, improves in Airline, and matches the baseline in Telecom.
With DeepSeek-V4-Flash, XFlow improves compliance in all three domains and improves or matches pass$^{1}$ in Airline and Telecom.
These results support the central claim that explicit protocol constraints can improve reliability without trading away task correctness.

\begin{table}[t]
\centering
\scriptsize
\setlength{\tabcolsep}{3pt}
\renewcommand{\arraystretch}{1.12}
\begin{tabular*}{\linewidth}{@{\extracolsep{\fill}}p{0.16\linewidth}p{0.20\linewidth}p{0.12\linewidth}rrr@{}}
\toprule
\textbf{Agent} & \textbf{Model} & \textbf{Domain} & \textbf{pass$^{1}$} & \textbf{Compliant} & \shortstack{\textbf{Conditional} \textbf{pass}} \\
\midrule
\rowcolor{blue!8}
\multicolumn{6}{c}{\textbf{$\tau^3$-bench}} \\
ReAct & Qwen3.5-9B & Retail & 65.5\% & 96.5\% & 63.2\% \\
+XFlow & Qwen3.5-9B & Retail & 63.2\% (\textcolor{red}{-2.3\%}) & 100.0\% (\textcolor{green!50!black}{+3.5\%}) & 63.2\% (\textcolor{black!55}{+0.0\%}) \\
ReAct & Qwen3.5-9B & Airline & 61.2\% & 91.8\% & 56.2\% \\
+XFlow & Qwen3.5-9B & Airline & 62.0\% (\textcolor{green!50!black}{+0.8\%}) & 100.0\% (\textcolor{green!50!black}{+8.2\%}) & 62.0\% (\textcolor{green!50!black}{+5.8\%}) \\
ReAct & Qwen3.5-9B & Telecom & 56.1\% & 77.2\% & 43.3\% \\
+XFlow & Qwen3.5-9B & Telecom & 56.1\% (\textcolor{black!55}{+0.0\%}) & 82.5\% (\textcolor{green!50!black}{+5.3\%}) & 46.3\% (\textcolor{green!50!black}{+3.0\%}) \\
ReAct & DeepSeek-V4-Flash & Retail & 87.6\% & 95.6\% & 83.7\% \\
+XFlow & DeepSeek-V4-Flash & Retail & 85.8\% (\textcolor{red}{-1.8\%}) & 99.1\% (\textcolor{green!50!black}{+3.5\%}) & 85.0\% (\textcolor{green!50!black}{+1.3\%}) \\
ReAct & DeepSeek-V4-Flash & Airline & 81.6\% & 87.8\% & 71.6\% \\
+XFlow & DeepSeek-V4-Flash & Airline & 85.7\% (\textcolor{green!50!black}{+4.1\%}) & 93.9\% (\textcolor{green!50!black}{+6.1\%}) & 80.4\% (\textcolor{green!50!black}{+8.8\%}) \\
ReAct & DeepSeek-V4-Flash & Telecom & 99.1\% & 63.7\% & 63.1\% \\
+XFlow & DeepSeek-V4-Flash & Telecom & 100.0\% (\textcolor{green!50!black}{+0.9\%}) & 100.0\% (\textcolor{green!50!black}{+36.3\%}) & 100.0\% (\textcolor{green!50!black}{+36.9\%}) \\
\bottomrule
\end{tabular*}
\caption{$\tau^3$-bench results with Qwen3.5-9B and DeepSeek-V4-Flash. The baseline is a bare ReAct agent. ReAct+XFlow wraps the same agent with protocol constraints. Parenthesized values report percentage-point changes relative to the corresponding ReAct row.}
\label{tab:tau3-results}
\end{table}
\begin{figure}[t]
\centering
\begin{minipage}[t]{0.48\linewidth}
\textbf{\textcolor{red!70!black}{ReAct: passed, non-compliant}}
\begin{lstlisting}[style=plain,language={},basicstyle=\ttfamily\scriptsize\color{black},keywordstyle=\color{black},stringstyle=\color{black},commentstyle=\color{black},breaklines=true,columns=fullflexible,escapeinside={(*@}{@*)}]
[9]  Assistant: Confirm Mei's DOB is still 1989-12-13?
[10] User: Yes. No other changes needed. Just update the name.
(*@\textcolor{red!70!black}{[11] Assistant -> update\_reservation\_passengers(...)}@*)
     (*@\textcolor{red!70!black}{// DOB was clarified, but the mutation itself was not explicitly confirmed.}@*)
Outcome: pass^1=true, compliant=false
\end{lstlisting}
\end{minipage}\hfill
\begin{minipage}[t]{0.48\linewidth}
\textbf{\textcolor{green!50!black}{XFlow: passed, compliant}}
\begin{lstlisting}[style=plain,language={},basicstyle=\ttfamily\scriptsize\color{black},keywordstyle=\color{black},stringstyle=\color{black},commentstyle=\color{black},breaklines=true,columns=fullflexible,escapeinside={(*@}{@*)}]
[9]  Assistant: Confirm Mei's DOB is still 1989-12-13?
[10] User: I do not have that handy; I just need the name changed.
(*@\textcolor{green!50!black}{[11] Assistant: I will keep the existing DOB. Please confirm that this is correct and that you want me to proceed.}@*)
[12] User: Yes, please go ahead and change it to Mei Garcia.
[13] Assistant -> update_reservation_passengers(...)
Outcome: pass^1=true, compliant=true
\end{lstlisting}
\end{minipage}
\caption{A matched $\tau^3$-bench trace where both agents solve the task, but only XFlow satisfies the confirmation-before-mutation constraint. XFlow delays the mutating tool call until the user has explicitly confirmed the action, not merely one of its parameters.}
\label{fig:tau3-trace-case}
\end{figure}
\paragraph{Case study.}
To understand the compliance gain beyond aggregate scores, we inspected matched trajectories.
In these cases, the ReAct baseline reached the correct end state but violated a process constraint.
XFlow both passed and remained compliant.
The pattern is not that the baseline fails to solve the user's task.
Rather, it often reaches the right answer through an invalid path.
The Airline passenger-name-change case in Figure~\ref{fig:tau3-trace-case} illustrates this distinction.
Both runs update ``Mei Lee'' to ``Mei Garcia'' and satisfy pass$^{1}$.
The difference is that the baseline treats a field clarification as permission to mutate the reservation, whereas XFlow keeps the mutation behind an explicit confirmation boundary.

This example illustrates XFlow's role in reliable agent execution.
The baseline reaches the correct final state, but it treats a parameter clarification as authorization for a state-changing action.
XFlow changes the execution path rather than the task goal.
Sensitive actions pass through protocol-visible gates, and mutation tools can be blocked or nudged until the required confirmation, disclosure, or obligation becomes an explicit event in the trace.

\subsection{Long-Context Reasoning: CorpusQA}
\label{sec:long-context-analysis}

CorpusQA evaluates whether an agent can preserve evidence visibility and answer commitment over long contexts.
These tasks stress a failure mode different from Constrained Interaction and Agentic Software Engineering.
The model may see a large amount of text, but the workflow still needs a structured account of selected evidence, aggregated claims, and committed answers.

The CorpusQA rows in Table~\ref{tab:corpusqa-agentic-results} report the long-context results.
The raw LLM rows measure direct model performance, while XpandA~\citep{xiao2025xpanda} rows add a long-context retrieval and aggregation baseline.
XpandA+XFlow is therefore a composition.
It keeps XpandA's long-context expansion, retrieval, and aggregation behavior, and adds XFlow's protocol features on top.
For chunk-level extraction over long documents, XFlow can declare contracts for expected fields, evidence spans, normalization rules, and validity predicates.
Extracted fragments then become checkable intermediate records before they are aggregated into a final answer.

\paragraph{Case study.}
Long-context financial questions often require more than retrieving the right table entries.
The agent must read accounting tables, extract the relevant quantities, and then apply domain-specific processing rules before making a comparison.
In these cases, the failure point is often not the raw value itself, but the rule used to interpret that value.
Figure~\ref{fig:corpusqa-financial-case} shows such a CorpusQA example.
The question defines ``cash used in investing activities'' as the absolute value of negative net cash from investing activities; positive investing cash flow denotes cash provided rather than cash used.
XpandA retrieves the relevant Texas Instruments values but compares the raw positive investing value against operating cash flow.
This produces a plausible but invalid answer.
XFlow instead names the financial quantities as symbols and derives \texttt{cash\_used} only when the investing value is negative.
It then stores a row-level \texttt{passes} predicate and assembles the final list only from rows satisfying that predicate.
The resulting answer matches the gold companies up to name normalization.

This example illustrates a broader point about protocol structure in long-context reasoning.
Some processing rules are too specific, numerous, or fragile to rely on as implicit model knowledge.
Even if they are written into a prompt, the agent may not consistently apply them during retrieval, aggregation, and answer formatting.
In XFlow, such requirements can be represented as derive rules over extracted symbols.
The model remains responsible for semantic reading, while the protocol performs rule-governed transformation and filtering before the final answer is committed.

\begin{table}[t]
\centering
\small
\setlength{\tabcolsep}{4pt}
\renewcommand{\arraystretch}{1.12}
\begin{tabular*}{\linewidth}{@{\extracolsep{\fill}}p{0.36\linewidth}p{0.36\linewidth}r@{}}
\toprule
\textbf{Agent} & \textbf{Model} & \textbf{Score} \\
\midrule
\rowcolor{blue!8}
\multicolumn{3}{c}{\textbf{CorpusQA}} \\
LLM & Qwen3.5-9B & 54.7 \\
XpandA & Qwen3.5-9B & 59.3 \\
+XFlow & Qwen3.5-9B & 61.7 (\textcolor{green!50!black}{+2.4}) \\
LLM & DeepSeek-V4-Flash & 57.8 \\
XpandA & DeepSeek-V4-Flash & 74.8 \\
+XFlow & DeepSeek-V4-Flash & 75.7 (\textcolor{green!50!black}{+0.9}) \\
\midrule
\rowcolor{blue!8}
\multicolumn{3}{c}{\textbf{SWE-bench Verified}} \\
mini-SWE-agent & DeepSeek-V4-Flash & 77.4 \\
+XFlow & DeepSeek-V4-Flash & 79.8 (\textcolor{green!50!black}{+2.4})\\
\bottomrule
\end{tabular*}
\caption{Combined Long-Context Reasoning and Agentic Software Engineering results with Qwen3.5-9B and DeepSeek-V4-Flash. Scores are accuracy for CorpusQA and pass rates for SWE-bench Verified. Parenthesized values report changes relative to the corresponding row for the same model.}
\label{tab:corpusqa-agentic-results}
\end{table}

\begin{figure}[t]
\centering
\begin{minipage}[t]{0.48\linewidth}
\textbf{\textcolor{red!70!black}{XpandA: raw comparison}}
\begin{lstlisting}[style=plain,basicstyle=\ttfamily\scriptsize,breaklines=true,columns=fullflexible]
Question: cash used in investing > operating

Texas Instruments:
  investing = 1253
  operating = 849
  decision: 1253 > 849

Answer:
  [GXO Logistics, Inc.,
   Texas Instruments]

Error: positive investing cash flow is cash provided, not cash used.
\end{lstlisting}
\end{minipage}\hfill
\begin{minipage}[t]{0.48\linewidth}
\textbf{\textcolor{green!50!black}{XFlow: derived constraint}}
\begin{lstlisting}[style=plain,basicstyle=\ttfamily\scriptsize,breaklines=true,columns=fullflexible]
symbols:
  investing, operating, cash_used, passes

derive:
  cash_used = abs(investing) if investing < 0 else None
  passes = cash_used is not None and cash_used > operating

rows:
  GXO:        -77,     29, true
  Ingredion:  -90,     77, true
  Silgan:  -82472, -683401, true
  Texas:     1253,    849, false

answer:
  [GXO Logistics, Inc., Ingredion Incorporated, 
  Silgan Holdings Inc.]
\end{lstlisting}
\end{minipage}
\caption{A CorpusQA financial example where XFlow turns a domain phrase into an executable derivation. XpandA retrieves relevant quantities but treats positive investing cash flow as cash used. XFlow applies the sign convention before answer assembly.}
\label{fig:corpusqa-financial-case}
\end{figure}

\subsection{Agentic Software Engineering: SWE-bench Verified}
\label{sec:agentic-task-analysis}

SWE-bench Verified evaluates long-horizon Agentic Software Engineering tasks.
An agent must inspect repository state, use tools over many turns, edit code, and produce a patch that an external evaluator later checks.
Compared with the preceding settings, which emphasize cooperation under constraints and evidence preservation over long documents, this setting stresses the reliability of extended tool interaction.
To maintain patch quality, the agent must follow basic software-engineering norms, such as running relevant tests or local reproducers.

These requirements can be stated in a prompt, but prompt instructions alone do not ensure that the agent will follow them throughout a long trajectory.
The Agentic Software Engineering rows in Table~\ref{tab:corpusqa-agentic-results} therefore evaluate XFlow as protocol structure around mini-SWE-agent~\citep{yang2024sweagent}.
The goal is not to change the primary actor model.
Instead, XFlow constrains long software-engineering workflows and enforces general process norms through explicit state, validation, and event records.

\paragraph{Case study.}
In \texttt{django\_\_django-11790}, the baseline mini-SWE-agent localized the issue to Django's \texttt{AuthenticationForm} and edited the username widget.
However, the submitted patch converted the widget's \texttt{maxlength} value to a string.
In our XPF protocol package, the submitted diff and the visible command trajectory are represented as protocol symbols.
Derived rules then compute process-level validation symbols.
They check whether the submission is a valid source-code diff.
They also check whether the final submission follows a passing local validation or reproducer.
In this case, the first attempt produced a structurally valid patch.
However, the derived process check remained false because the agent submitted after a local validation command still failed.
The protocol therefore did not accept the submission handoff and instead produced a targeted retry signal.
On the retry, the agent preserved the integer \texttt{max\_length} value.
It then ran a focused \texttt{AuthenticationFormTest} check successfully and submitted a patch that was later resolved by the official SWE-bench evaluator.

\begin{figure}[t]
\centering
\begin{minipage}[t]{0.48\linewidth}
\textbf{\textcolor{red!70!black}{mini-SWE-agent: submits after failed validation}}
\begin{lstlisting}[style=plain,basicstyle=\ttfamily\scriptsize,breaklines=true,columns=fullflexible]
attempt 1:
  edit AuthenticationForm at the right site
  maxlength = str(max_length)

local validation:
  tests/runtests.py auth_tests -> FAILED

submission decision:
  baseline process still submits patch

official result:
  target tests still failing
\end{lstlisting}
\end{minipage}\hfill
\begin{minipage}[t]{0.48\linewidth}
\textbf{\textcolor{green!50!black}{XFlow: submit gated by passing validation}}
\begin{lstlisting}[style=plain,basicstyle=\ttfamily\scriptsize,breaklines=true,columns=fullflexible,escapeinside={(*@}{@*)}]
attempt 1:
  edit AuthenticationForm at the right site
  maxlength = str(max_length)

local validation:
  tests/runtests.py auth_tests -> FAILED
  (*@\textcolor{xpfKeyword}{XFlow derive rules are triggered and return feedback.}@*)
  (*@\textcolor{xpfKeyword}{The message asks the agent to keep revising.}@*)

attempt 2:
  maxlength = max_length
  preserve integer field value

submission decision:
  gate opens only after passing check

official result:
  target tests pass
\end{lstlisting}
\end{minipage}
\caption{A SWE-bench Verified case where process-derived feedback recovers a near-miss patch. Without XFlow, the agent submits despite a failed local validation command. With XFlow, submission is gated on a passing check. The retry preserves the integer \texttt{maxlength} value.}
\label{fig:swebench-django-11790-case}
\end{figure}
\FloatBarrier

\begin{takeaway}
The experiments illustrate three uses of structured knowledge.
In $\tau^3$-bench, task constraints can be externalized as checkable rules.
In long-context reasoning, domain facts and interpretation rules can be encoded as deterministic derivations.
In Agentic Software Engineering, procedures can be represented as stages, loops, gates, and event records.
Across these settings, moving reusable constraints, facts, and workflows from actor memory into protocol-governed structure makes agent systems more reliable.
\end{takeaway}

\section{Use Cases in Cloud--Edge Agent Systems}
\label{sec:cloud_edge}

Beyond benchmark evaluation, XFlow is applicable to deployment settings where cloud-side planning must coordinate with resource-constrained edge-side execution.
The core challenge is not only division of labor, but also protocol governance.
Edge models should be useful without being allowed to redefine the task, invoke tools at arbitrary times, or propagate uncertain intermediate outputs into global state.
XFlow makes this cloud--edge boundary executable.
It declares what each side may see, produce, and commit.
It also declares when cloud and edge actors may interact, and which validation or review events must happen before uncertain local results become shared workflow state.
We illustrate this application setting with two case studies: Cloud--Edge Long-Context Reasoning and DeepResearch.

\subsection{Cloud--Edge Long-Context Reasoning}

Long-document question answering and analysis expose a common failure mode of
prompt-driven multi-agent systems.
The task is semantically open: the system must read a large input, locate evidence,
compare dispersed information, and synthesize a final answer.
Yet the execution structure should not be open-ended.
A planner may decompose the document inconsistently.
A worker may reason outside its assigned scope.
Missing evidence may hide inside a fluent answer, and evaluator feedback may remain only a prompt-level suggestion.
Cloud--Edge Long-Context Reasoning therefore stresses the formal/informal boundary that motivates XFlow.
Semantic reasoning remains flexible, but interfaces, visibility, evidence requirements, and revision decisions must be governed by protocol.

\begin{figure}[t]
    \centering
    \includegraphics[width=\textwidth]{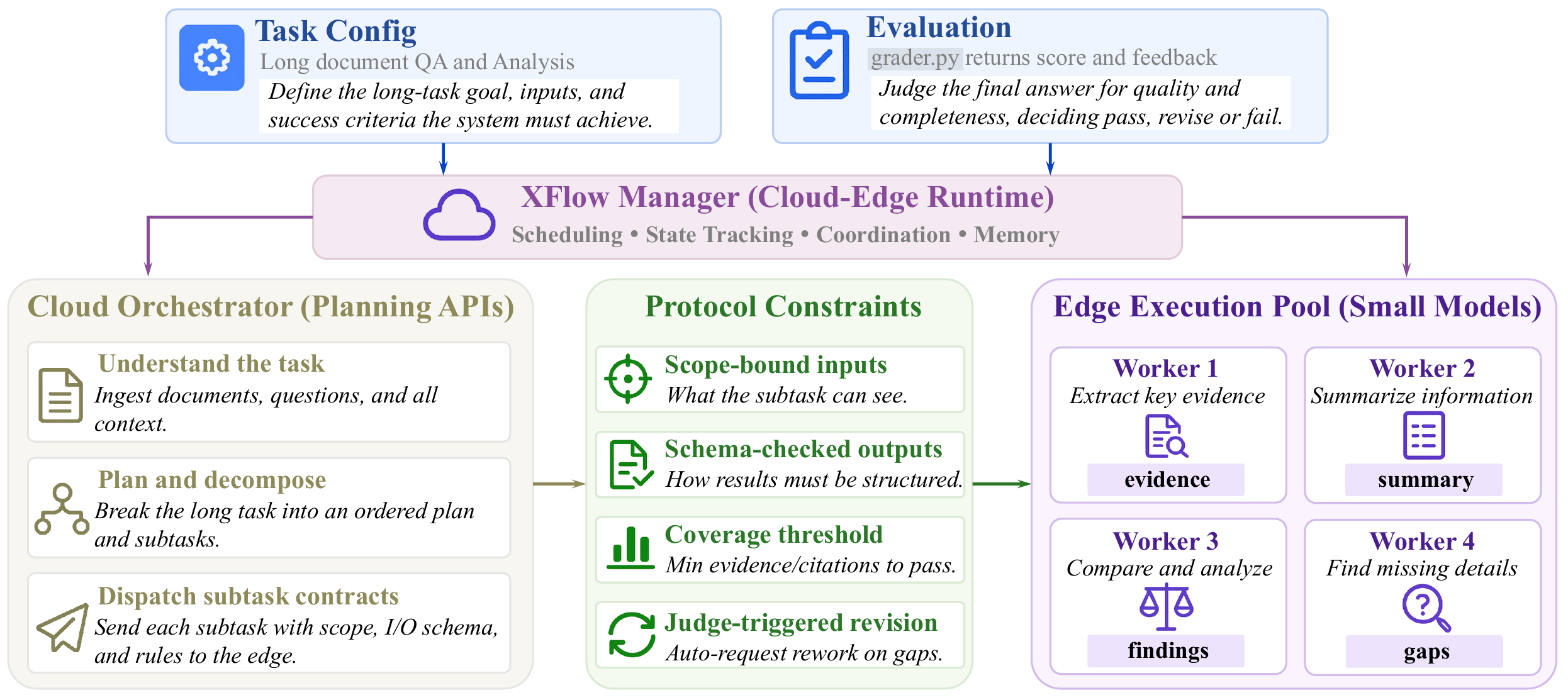}
    \caption{
    \textbf{Long-context execution as a cloud--edge use case.}
    XFlow separates global planning from bounded local execution. A cloud-side
    orchestrator ingests the task, decomposes the long document, and dispatches scoped
    subtask contracts. Edge-side small models then perform extraction, summarization,
    comparison, and gap finding under declared constraints.
    }
    \label{fig:longtask_case}
\end{figure}

Figure~\ref{fig:longtask_case} illustrates this boundary in a cloud--edge setting. The
cloud-side orchestrator handles global decisions that require broad context, including task
understanding, document segmentation, planning, and subtask construction. Each generated
subtask is packaged as a contract with a declared scope, input/output schema, local rules,
and continuation point. The edge execution pool then runs smaller worker models on these
bounded contracts. Workers do not receive the whole workflow state. They read only the
assigned chunks and write only declared outputs such as evidence, summaries, findings, or
gap reports. Their outputs must pass schema validation and coverage checks before they
enter shared state. This prevents local uncertainty from silently becoming part of the
global answer context.

The protocol makes decomposition and aggregation inspectable. The orchestrator's plan is
treated as a proposed value rather than immediately committed state. Derived checks expose
whether chunks, subtasks, assignments, and coverage are sufficient for execution. After
parallel edge execution, XFlow lifts branch-local results back into global state. It groups
findings, preserves source references, and makes missing coverage explicit. The
synthesizer therefore writes the final answer from a structured evidence base rather than
from an unordered transcript of agent messages.

Evaluation is also represented as a protocol boundary. The grader returns a score and
feedback on answer quality and completeness. XFlow converts this feedback into
declared judge symbols. Flow rules then decide whether the answer passes, enters targeted
revision, or fails. During revision, the reviser receives the previous answer, structured
evidence, and grader feedback. Repair is therefore triggered by explicit evaluation state
rather than by an informal instruction to ``try again.'' If the answer passes, the publish
stage emits the final response and a manifest that records quality and coverage status.
If the revision budget is exhausted, XFlow can still publish a best-effort answer.
The manifest then records the remaining quality and coverage status.

This application demonstrates XFlow as a cloud--edge runtime boundary.
Planning, extraction, summarization, comparison, and grading remain semantic operations performed by LLM actors, APIs, or small edge models.
XFlow formalizes the structure around them.
It declares which inputs each actor may see, what outputs it may write, and when cloud and edge actors may interact.
It also declares when intermediate results become shared state, how grader feedback changes control flow, and what evidence status accompanies the final answer.
Cloud--Edge Long-Context Reasoning is therefore not merely a multi-agent prompt chain over a long document.
It is an executable protocol for constraining edge behavior, regulating interaction timing, bounding uncertainty propagation, and coordinating judge-triggered revision.

\subsection{DeepResearch: Protocolizing Cloud--Edge Interaction}

DeepResearch is the open-ended counterpart to the Cloud--Edge Long-Context Reasoning use case.
Instead of a fixed long-context input, the system must plan a research strategy, run web searches, and extract evidence.
It must also synthesize findings, draft a report, review it, revise deficiencies, and format the final answer.
The semantic space is broad, but the execution still needs bounded structure.
Edge workers should not rewrite the research question, expand tool calls arbitrarily, introduce unsupported claims, or commit unverified outputs into the report.
XFlow therefore turns an exploratory research agent into a protocol-governed cloud--edge workflow with explicit interaction points and uncertainty boundaries.
The workflow adopts a cloud--edge division of labor.
Cloud-side actors handle global semantic decisions, including search planning, synthesis, report writing, quality and compliance judging, revision, and final formatting.
Edge-side workers are assigned only local executable actions.
\texttt{web\_searcher} runs one read-only sub-query.
\texttt{content\_extractor} fetches selected pages.
\texttt{fact\_extractor} converts synthesized text into structured claims with sources, evidence, and confidence scores.
These workers are not autonomous research agents.
They operate inside declared inputs, outputs, tool permissions, continuation points, and validation gates, so their responses cannot silently redefine workflow state or advance the interaction before the protocol allows it.

\begin{figure}[t]
    \centering
    \includegraphics[width=\textwidth]{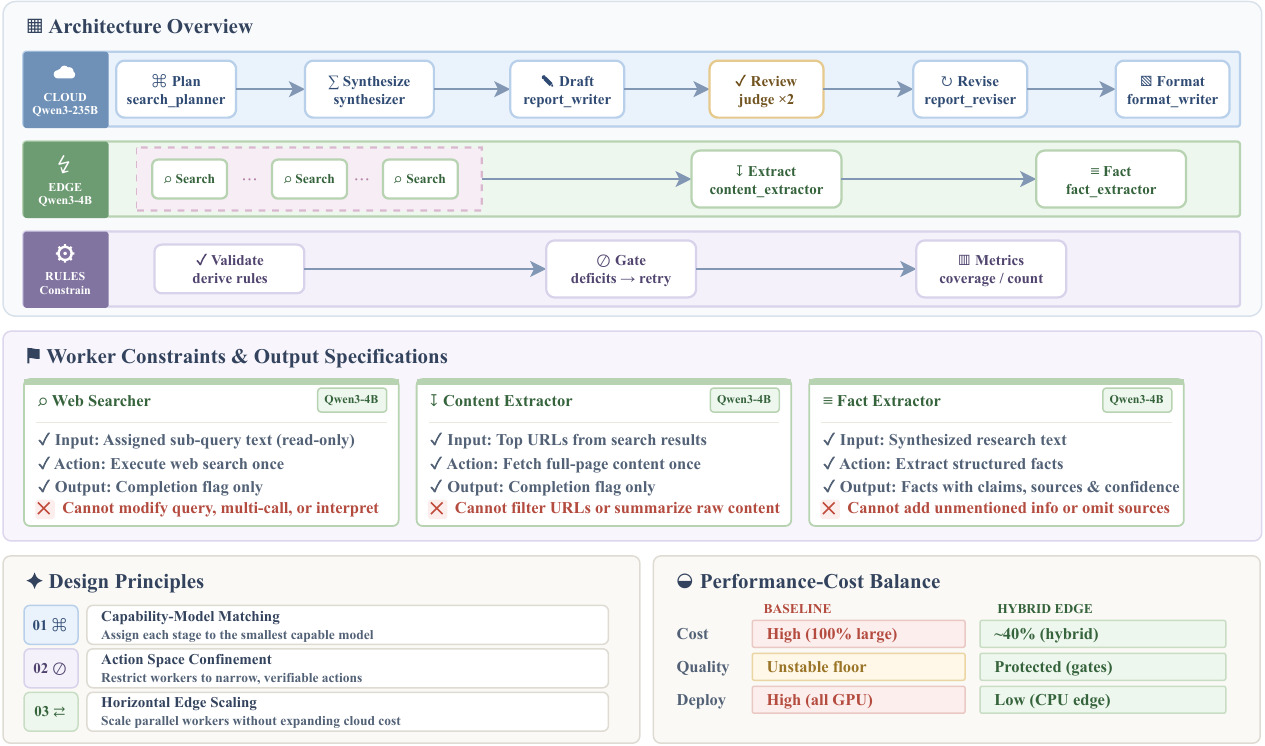}
    \caption{
    \textbf{DeepResearch as protocolized cloud--edge interaction.}
    XFlow assigns global research decisions to cloud-side actors while confining
    edge workers to narrow search, extraction, and fact-extraction actions. The
    protocol records artifacts, validates derived research sufficiency metrics,
    and gates cloud--edge continuation, retry, or revision before a final report
    is formatted for delivery.
    }
    \label{fig:deepresearch_case}
\end{figure}

XFlow makes this constrained action space inspectable.
The protocol divides research into stages such as planning, parallel search, extraction, synthesis, fact extraction, validation, drafting, review, revision, and finalization.
These stages define what each actor does and when cloud-side and edge-side actors may exchange state.
Derived checks expose whether planning and evidence are sufficient: sub-query count, source diversity, extraction success, fact coverage, confidence statistics, report structure, and citation coverage.
Thus, vague instructions such as ``conduct sufficient research'' become executable conditions.
They can trigger retry, revision, or best-effort delivery rather than leaving evidence gaps hidden in a fluent final report.

The same protocol preserves continuity across long, branching execution.
Search results, extracted pages, synthesized findings, structured facts, draft reports, review verdicts, and revision status are represented as symbols, derived symbols, or artifacts.
Here, artifacts mean files or records produced during execution.
A handoff exposes committed state to the next stage rather than an informal chat transcript.
Unvalidated artifacts remain local until they cross declared checks.
The \texttt{synthesizer} receives structured search and extraction aggregates.
The \texttt{report\_reviser} receives explicit deficits such as \texttt{quality\_reason}, \texttt{compliance\_issues}, and \texttt{draft\_deficits}.
Because tool outputs are stored as artifacts, later stages can reason from the same evidence record without requiring small models to copy raw pages into free-form responses.

This application demonstrates XFlow's main cloud--edge advantage: it does not require every actor to be globally reliable.
Instead, it makes small edge models locally useful by restricting their authority and validating their outputs.
It also regulates when they may interact with cloud-side state and prevents uncertain local artifacts from propagating unchecked.
DeepResearch is therefore not simply a multi-agent search pipeline.
It is an executable research protocol that coordinates open-ended actors through reliable boundaries.

\FloatBarrier

\begin{takeaway}
The semi-formal XPF protocol is essential for cloud--edge collaboration.
Edge-side models cannot be assumed to execute tasks with perfect reliability, so XFlow constrains model behavior within declared protocol boundaries.
It also uses symbols to capture unreliable situations as explicit workflow state.
This makes uncertainty visible and controllable rather than allowing it to silently propagate through the workflow.
\end{takeaway}

\section{Related Work}
\label{sec:related_work}

\subsection{Agent Orchestration Frameworks}

Recent LLM application frameworks span a broad orchestration design space.
At one end, \textbf{LangChain}~\citep{langchain} and \textbf{LlamaIndex}~\citep{llamaindex} popularized component-level abstractions for chains, retrieval, tools, memory, and agent loops.
These abstractions make it easy to compose LLM applications from reusable modules.
\textbf{LangGraph}~\citep{langgraph} moves this space toward explicit graph runtimes for stateful, long-running agents.
It provides nodes, edges, checkpointing, durable execution, human-in-the-loop interruption, and multi-agent patterns.
Conversation-centered systems such as \textbf{AutoGen}~\citep{autogen} organize applications as programmable interactions among conversable agents.
Role- and task-based frameworks such as \textbf{CrewAI}~\citep{crewai} expose teams, tasks, crews, and flows as higher-level orchestration objects.

Research prototypes further explore multi-agent coordination as social or organizational structure.
\textbf{CAMEL}~\citep{li2023camel} uses role-playing communicative agents.
\textbf{MetaGPT}~\citep{hong2023metagpt} encodes software-development standard operating procedures.
\textbf{ChatDev}~\citep{qian2024chatdev} implements a chat-powered virtual software company.
\textbf{AgentVerse}~\citep{chen2024agentverse} studies dynamically composed multi-agent groups and emergent collaboration.
Recent specification-oriented systems move closer to XFlow.
\textbf{AgentSPEX}~\citep{wang2026agentspex} specifies LLM-agent workflows with explicit control flow, modular structure, typed steps, branching and loops.
It also supports parallel execution, reusable submodules, explicit state management, and a customizable execution harness.

These systems demonstrate the practical value of agent orchestration, but they often leave key workflow commitments embedded in library code, graph callbacks, conversation policies, or workflow configuration.
Examples include symbol ownership, visibility, typed state, commit conditions, recovery paths, and allowable transitions.
This makes the prompt--harness boundary difficult to inspect, validate, replay, or enforce independently of runtime execution.
XFlow addresses this gap by treating orchestration as a literate executable protocol.
XPF is compiled into a typed IR.
The runtime then enforces lifecycle-governed symbols, commit policies, handoff gates, event-sourced replay, and recovery semantics.

\subsection{Reliability in LLMs and Agents}

Reliability has become a central concern for LLMs and LLM-based agents because errors are not limited to wrong final answers.
Hallucination studies show that fluent outputs can still contain unsupported claims~\citep{ji2023survey}, and Xu et al.~\citep{xu2025delusions} show that high-confidence hallucinations are especially hard to detect.
This motivates work on knowledge boundaries.
Xu et al.~\citep{xu2024rejection} train models to refuse out-of-scope questions, Zheng et al.~\citep{zheng2025ekbm} make the boundary explicit, and earlier self-knowledge work asks whether LLMs can recognize what they do not know~\citep{yin2023selfknowledge}.
These studies shift reliability from answer generation alone toward knowing when an answer should be withheld or treated with caution.
In agents, the same issue extends to action.
Tool-use benchmarks and training frameworks such as ToolLLM~\citep{qin2024toolllm} expand API-calling ability.
Xu et al.~\citep{xu2024toolhallucination,xu2025efficienttoolcalling} study how to reduce tool hallucination and unnecessary tool calls.
Routing work makes a related point.
Ong et al.~\citep{ong2024routellm} learn centralized quality--cost routers, and Zheng et al.~\citep{zheng2026disrouter} study distributed self-routing among LLM agents.
Together, these works show that reliable agents must judge not only what to say, but also when to refuse, call a tool, or pass responsibility to another model.

These works mainly improve model-side judgment, refusal, tool selection, or routing.
XFlow addresses a complementary workflow-side problem.
Even when actors remain uncertain, the system must decide when their outputs can become shared state, when a tool call is allowed, and who should act next.
XFlow externalizes these decisions into executable protocol state.
Actors may remain semantically open, while interfaces, state transitions, commit gates, handoffs, recovery, and audit records become visible protocol structure.

\section{Conclusion}
\label{sec:conclusion}

We presented \textbf{XFlow} as an executable protocol programming system for reliable multi-agent workflows.
The central argument is that reliability depends on where a workflow draws the formal/informal boundary.
XPF makes this boundary explicit through semi-formal protocol structure.
Protocol authors can keep informal semantic work inside LLM and human actors while exposing tools through declared contracts.
They can externalize selected commitments as symbols, policies, flow rules, and lifecycle transitions.
The runtime can then inspect, preserve, and enforce those commitments.
This design prevents uncertain actor outputs from becoming shared state until they pass declared protocol boundaries.
Across Constrained Interaction, Long-Context Reasoning, Agentic Software Engineering, and cloud--edge use cases, XFlow shows how externalizing reusable constraints and process commitments can improve reliability without removing actor flexibility.
Overall, XFlow argues for treating the prompt--harness boundary as a first-class abstraction for building multi-agent systems that are easier to inspect, recover, and trust.

\subsection*{Limitations and Future Work}

XPF requires protocol authors to think in terms of stages, symbols, and policies, which introduces a learning curve compared with writing prompts directly.
Designing a well-structured protocol requires upfront work to decompose a workflow into stages with explicit interfaces.
This decomposition may not always be straightforward for tasks that are fluid or exploratory.
Future work includes richer tooling for protocol authoring, such as interactive templates, visual editors, and AI-assisted scaffolding that can lower the barrier to adoption.
XFlow also does not provide formal guarantees about the semantic quality of actor outputs.
Schema validation verifies shape, but not factual correctness or logical soundness.
Future work may explore contract-based validation, where actors declare postconditions that can be partially checked by tools or weaker models.
This would provide a middle ground between schema validation and full semantic verification.

\clearpage

\bibliography{references}

@article{gray1981transaction,
  title={The Transaction Concept: Virtues and Limitations},
  author={Gray, Jim},
  journal={Proceedings of the Seventh International Conference on Very Data Bases},
  pages={144--154},
  year={1981}
}

@article{chomsky1956three,
  title={Three Models for the Description of Language},
  author={Chomsky, Noam},
  journal={IRE Transactions on Information Theory},
  volume={2},
  number={3},
  pages={113--124},
  year={1956},
  publisher={IEEE}
}

@article{park2023generative,
  title = {Generative Agents: Interactive Simulacra of Human Behavior},
  author = {Joon Sung Park and Joseph C. O'Brien and Carrie Jun Cai and Meredith Ringel Morris and Percy Liang and Michael S. Bernstein},
  year = {2023},
  journal = {UIST}
}

@article{hong2023metagpt,
  title = {MetaGPT: Meta Programming for A Multi-Agent Collaborative Framework},
  author = {Sirui Hong and Mingchen Zhuge and Jiaqi Chen and Xiawu Zheng and Yuheng Cheng and Jinlin Wang and Ceyao Zhang and Zili Wang and Steven Ka Shing Yau and Zijuan Lin and Liyang Zhou and Chenyu Ran and Lingfeng Xiao and Chenglin Wu and Jürgen Schmidhuber},
  year = {2023},
  journal = {ICLR}
}

@article{yao2023react,
  title={{ReAct}: Synergizing Reasoning and Acting in Language Models},
  author={Yao, Shunyu and Zhao, Jeffrey and Yu, Dian and Du, Nan and Shafran, Izhak and Narasimhan, Karthik and Cao, Yuan},
  journal={International Conference on Learning Representations (ICLR)},
  year={2023}
}

@article{bainomugisha2013survey,
  title={A Survey on Reactive Programming},
  author={Bainomugisha, Engineer and Carreton, Andoni Lombide and Cutsem, Tom Van and Mostinckx, Stijn and Meuter, Wolfgang De},
  journal={ACM Computing Surveys (CSUR)},
  volume={45},
  number={4},
  pages={1--34},
  year={2013},
  publisher={ACM}
}

@inproceedings{langchain,
  title = {Creating large language model applications utilizing langchain: A primer on developing llm apps fast},
  author = {O Topsakal and TC Akinci},
  year = {2023},
  booktitle = {International conference on applied ...}
}

@misc{llamaindex,
  author    = {Jerry Liu},
  title     = {LlamaIndex: A data framework for LLM applications},
  year      = {2023},
  url       = {https://github.com/run-llama/llama_index},
  note      = {Open source framework}
}

@article{autogen,
  title = {AutoGen: Enabling Next-Gen LLM Applications via Multi-Agent Conversation},
  author = {Qingyun Wu and Gagan Bansal and Jieyu Zhang and Yiran Wu and Beibin Li and Erkang Zhu and Li Jiang and Xiaoyun Zhang and Shaokun Zhang and Jiale Liu and Ahmed Hassan Awadallah and Ryen W White and Doug Burger and Chi Wang},
  year = {2023},
  journal = {arXiv (Cornell University)}
}

@misc{crewai,
  author    = {Jo{\~a}o Moura},
  title     = {CrewAI: Framework for orchestrating role-playing, autonomous AI agents},
  year      = {2024},
  url       = {https://github.com/joaomdmoura/crewAI},
  note      = {Open source framework}
}

@article{wei2022chain,
  title = {Chain of Thought Prompting Elicits Reasoning in Large Language Models},
  author = {Jason Wei and Xuezhi Wang and Dale Schuurmans and Maarten Bosma and Ed H. Chi and Denny Zhou},
  year = {2022},
  journal = {Neural Information Processing Systems}
}

@article{xi2025rise,
  title = {The Rise and Potential of Large Language Model Based Agents: A Survey},
  author = {Zhiheng Xi and Wenxiang Chen and Xin Guo and Wei He and Yiwen Ding and Boyang Hong and Ming Zhang and Junzhe Wang and Senjie Jin and Enyu Zhou and Rui Zheng and Xiaoran Fan and Xiao Wang and Limao Xiong and Yuhao Zhou and Weiran Wang and Changhao Jiang and Yicheng Zou and Xiangyang Liu and Zhangyue Yin and Shihan Dou and Rongxiang Weng and Wensen Cheng and Qi Zhang and Wenjuan Qin and Yongyan Zheng and Xipeng Qiu and Xuanjing Huang and Tao Gui},
  year = {2023},
  journal = {arXiv (Cornell University)}
}

@article{ji2023survey,
  title = {Survey of Hallucination in Natural Language Generation},
  author = {Ziwei Ji and Nayeon Lee and Rita Frieske and Tiezheng Yu and Dan Su and Yan Xu and Etsuko Ishii and Yejin Bang and Andrea Madotto and Pascale Fung},
  year = {2022},
  journal = {ACM Comput. Surv.}
}

@inproceedings{elliott1997functional,
  title={Functional Reactive Animation},
  author={Elliott, Conal and Hudak, Paul},
  booktitle={Proceedings of the 2nd ACM SIGPLAN International Conference on Functional Programming},
  pages={263--273},
  year={1997}
}

@article{li2023camel,
  title={{CAMEL}: Communicative Agents for Mind Exploration of Large Scale Language Model Society},
  author={Li, Guohao and Hammoud, Hasan Abed Al Kader and Itani, Hadi and Khizbullin, Dmitrii and Ghanem, Bernard},
  journal={arXiv preprint arXiv:2303.17760},
  year={2023}
}

@inproceedings{yao2024taubench,
  title = {{$\tau$-bench}: A Benchmark for Tool-Agent-User Interaction in Real-World Domains},
  author = {Yao, Shunyu and Shinn, Noah and Razavi, Pedram and Narasimhan, Karthik},
  booktitle = {International Conference on Learning Representations},
  year = {2025}
}

@article{barres2025tau2,
  title = {{$\tau^2$-Bench}: Evaluating Conversational Agents in a Dual-Control Environment},
  author = {Barres, Victor and Dong, Honghua and Ray, Soham and Si, Xujie and Narasimhan, Karthik},
  journal = {arXiv preprint arXiv:2506.07982},
  year = {2025}
}

@article{lu2026corpusqa,
  title = {{CorpusQA}: A 10 Million Token Benchmark for Corpus-Level Analysis and Reasoning},
  author = {Lu, Zhiyuan and Li, Chenliang and Shi, Yingcheng and Shen, Weizhou and Yan, Ming and Huang, Fei},
  journal = {arXiv preprint arXiv:2601.14952},
  year = {2026}
}

@inproceedings{jimenez2024swebench,
  title = {{SWE-bench}: Can Language Models Resolve Real-World {GitHub} Issues?},
  author = {Jimenez, Carlos E. and Yang, John and Wettig, Alexander and Yao, Shunyu and Pei, Kexin and Press, Ofir and Narasimhan, Karthik},
  booktitle = {International Conference on Learning Representations},
  year = {2024}
}

@misc{qwen2026qwen35,
  title = {{Qwen3.5-9B}},
  author = {{Qwen Team}},
  year = {2026},
  url = {https://huggingface.co/Qwen/Qwen3.5-9B},
  note = {Model card}
}

@misc{deepseekai2026deepseekv4,
  title = {{DeepSeek-V4-Flash}},
  author = {{DeepSeek-AI}},
  year = {2026},
  url = {https://huggingface.co/deepseek-ai/DeepSeek-V4-Flash},
  note = {Model card}
}

@article{xiao2025xpanda,
  title = {Long Context Scaling: Divide and Conquer via Multi-Agent Question-Driven Collaboration},
  author = {Xiao, Sibo and Lin, Zixin and Gao, Wenyang and Zhang, Yue},
  journal = {arXiv preprint arXiv:2505.20625},
  year = {2025}
}

@article{yang2024sweagent,
  title = {SWE-agent: Agent-Computer Interfaces Enable Automated Software Engineering},
  author = {Carlos Jimenez and Kilian Lieret and Karthik Narasimhan and Ofir Press and Alexander Wettig and John Yang and Shunyu Yao},
  year = {2024},
  journal = {Advances in Neural Information Processing Systems 37}
}

@misc{openai2024swebenchverified,
  title = {Introducing {SWE-bench Verified}},
  author = {{OpenAI}},
  year = {2024},
  url = {https://openai.com/index/introducing-swe-bench-verified/},
  note = {OpenAI blog}
}

@misc{langgraph,
  author = {{LangChain}},
  title = {{LangGraph}: Build Resilient Language Agents as Graphs},
  year = {2026},
  url = {https://docs.langchain.com/oss/python/langgraph},
  note = {Documentation}
}

@article{wang2026agentspex,
  title = {{AgentSPEX}: An Agent {SP}ecification and {EX}ecution Language},
  author = {Wang, Pengcheng and Huang, Jerry and Yao, Jiarui and Pan, Rui and Niu, Peizhi and Liu, Yaowenqi and Wang, Ruida and Lu, Renhao and Guo, Yuwei and Zhang, Tong},
  journal = {arXiv preprint arXiv:2604.13346},
  year = {2026},
  url = {https://arxiv.org/abs/2604.13346}
}

@inproceedings{qian2024chatdev,
  title = {{ChatDev}: Communicative Agents for Software Development},
  author = {Qian, Chen and Liu, Wei and Liu, Hongzhang and Chen, Nuo and Dang, Yufan and Li, Jiahao and Yang, Cheng and Chen, Weize and Su, Yusheng and Cong, Xin and Xu, Juyuan and Li, Dahai and Liu, Zhiyuan and Sun, Maosong},
  booktitle = {Proceedings of the 62nd Annual Meeting of the Association for Computational Linguistics},
  year = {2024}
}

@inproceedings{chen2024agentverse,
  title = {{AgentVerse}: Facilitating Multi-Agent Collaboration and Exploring Emergent Behaviors},
  author = {Chen, Weize and Su, Yusheng and Zuo, Jingwei and Yang, Cheng and Yuan, Chenfei and Chan, Chi-Min and Yu, Heyang and Lu, Yaxi and Hung, Yi-Hsin and Qian, Chen and Qin, Yujia and Cong, Xin and Xie, Ruobing and Liu, Zhiyuan and Sun, Maosong and Zhou, Jie},
  booktitle = {International Conference on Learning Representations},
  year = {2024}
}

@article{xu2025delusions,
  title = {Delusions of Large Language Models},
  author = {Xu, Hongshen and Yang, Zixv and Zhu, Zichen and Lan, Kunyao and Wang, Zihan and Wu, Mengyue and Ji, Ziwei and Chen, Lu and Fung, Pascale and Yu, Kai},
  journal = {arXiv preprint arXiv:2503.06709},
  year = {2025}
}

@inproceedings{xu2024rejection,
  title = {Rejection Improves Reliability: Training {LLMs} to Refuse Unknown Questions Using {RL} from Knowledge Feedback},
  author = {Xu, Hongshen and Zhu, Zichen and Zhang, Situo and Ma, Da and Fan, Shuai and Chen, Lu and Yu, Kai},
  booktitle = {Conference on Language Modeling},
  year = {2024}
}

@article{zheng2025ekbm,
  title = {Enhancing {LLM} Reliability via Explicit Knowledge Boundary Modeling},
  author = {Zheng, Hang and Xu, Hongshen and Liu, Yuncong and Chen, Lu and Fung, Pascale and Yu, Kai},
  journal = {arXiv preprint arXiv:2503.02233},
  year = {2025}
}

@article{yin2023selfknowledge,
  title = {Do Large Language Models Know What They Don't Know?},
  author = {Yin, Zhangyue and Sun, Qiushi and Guo, Qipeng and Wu, Jiawen and Qiu, Xipeng and Huang, Xuanjing},
  journal = {Findings of the Association for Computational Linguistics: ACL},
  year = {2023}
}

@article{qin2024toolllm,
  title = {ToolLLM: Facilitating Large Language Models to Master 16000+ Real-world APIs},
  author = {Yujia Qin and Shihao Liang and Yining Ye and Kunlun Zhu and Lan Yan and Yaxi Lu and Yankai Lin and Xin Cong and Xiangru Tang and Bill Qian and Sihan Zhao and Lauren Hong and Runchu Tian and Ruobing Xie and Jie Zhou and Mark Gerstein and Dahai Li and Zhiyuan Liu and Maosong Sun},
  year = {2023},
  journal = {ICLR}
}

@article{xu2024toolhallucination,
  title = {Reducing Tool Hallucination via Reliability Alignment},
  author = {Xu, Hongshen and Zhu, Zichen and Pan, Lei and Wang, Zihan and Zhu, Su and Ma, Da and Cao, Ruisheng and Chen, Lu and Yu, Kai},
  journal = {arXiv preprint arXiv:2412.04141},
  year = {2024}
}

@article{xu2025efficienttoolcalling,
  title = {Alignment for Efficient Tool Calling of Large Language Models},
  author = {Hongshen Xu and Zihan Wang and Zichen Zhu and Lei Pan and Xingyu Chen and Shuai Fan and Lu Chen and Kai Yu},
  year = {2025},
  journal = {Proceedings of the 2025 Conference on Empirical Methods in Natural Language Processing}
}

@article{ong2024routellm,
  title = {{RouteLLM}: Learning to Route {LLMs} with Preference Data},
  author = {Ong, Isaac and Almahairi, Amjad and Wu, Vincent and Chiang, Wei-Lin and Wu, Tianhao and Gonzalez, Joseph E. and Kadous, M. Waleed and Stoica, Ion},
  journal = {arXiv preprint arXiv:2406.18665},
  year = {2024}
}

@article{zheng2026disrouter,
  title = {{DiSRouter}: Distributed Self-Routing for {LLM} Selections},
  author = {Zheng, Hang and Xu, Hongshen and Lin, Yongkai and Fan, Shuai and Chen, Lu and Yu, Kai},
  journal = {arXiv preprint arXiv:2510.19208},
  year = {2025}
}
\bibliographystyle{xlance/IEEEtran2}

\clearpage
\appendix
\section{XPF Protocol Syntax Reference}
\label{app:xpf-syntax}

This appendix gives the syntax reference for XPF, the protocol authoring format used by XFlow.
It is organized from the package boundary inward.
A.1 describes the source package that the compiler loads.
A.2 describes the structure of the root \texttt{protocol.xpf} document.
A.3--A.6 define protocol headers, sidecar definition files, symbols, policies, and stages.
A.7--A.9 describe semantic blocks, predicates, calls, and handoffs.
A.10--A.11 collect well-formedness rules and a compact protocol slice.

\subsection{XPF Package Structure}

An XPF package is the source bundle that the compiler loads.
The minimal package contains a root \texttt{protocol.xpf} file.
Larger packages may add sidecar definitions for actors, profiles, tools, schemas, derivation code, and child protocols.
The source package describes the executable workflow contract.
Runtime products such as session logs, pending approvals, and generated artifacts are produced during execution and are not part of the source package itself.

\par\noindent\begin{minipage}{\linewidth}
\begin{lstlisting}[style=plain]
research-brief-demo/              # package root
  protocol.xpf
  agents/                         # actor sidecar definitions
    writer.agent.md
    reviewer.agent.md
  profiles/                       # invocation defaults
    drafting-base.profile.md
  tools/                          # callable tool contracts
    compose_brief.tool.md
  schemas/                        # reusable structured types
    brief.schema.yaml
  rules/                          # package-local deterministic code
    readiness.py
  revise-draft/                   # nested child protocol package
    protocol.xpf
\end{lstlisting}
\end{minipage}\par

The root \texttt{protocol.xpf} file provides the protocol header and stage document.
Files under \texttt{agents/} and \texttt{profiles/} provide reusable actor instructions and invocation defaults.
Files under \texttt{tools/} expose callable tool contracts, while \texttt{schemas/} stores reusable JSON-schema-shaped type definitions.
Files under \texttt{rules/} or another package-local module directory can implement deterministic helper functions for derived values.
A child directory with its own \texttt{protocol.xpf} is a nested protocol package.
The parent header can reference it through \texttt{subflows} and enter it through a \texttt{call} transition.

XPF uses explicit package-local references rather than hidden imports.
A \texttt{subflows} entry maps a local alias to a child package or child \texttt{protocol.xpf} file.
Actor, profile, tool, and schema names resolve to sidecar definitions by identifier.
A derive block may either use an inline \texttt{expr} or reference package code with a \texttt{python} field such as \texttt{rules.readiness:draft\_ready}.
These references are import-like because they name external files in the package.
They remain visible in the protocol text and are checked relative to the package root.

\subsection{Protocol Document Structure}

The root \texttt{protocol.xpf} document has three layers.
The YAML frontmatter is the \emph{protocol header}: it declares global objects such as the protocol identifier, entry stage, symbols, actors, and policies.
Markdown headings and prose form the \emph{literate stage document}: headings define the stage hierarchy, while prose remains readable context for humans and actors.
Fenced semantic blocks are the \emph{executable declarations}: they bind actors, declare reads and writes, define derivations, judges, handoffs, and flow transitions for the surrounding stage.
The code example below mirrors these layers.
The header names the protocol, entry stage, state symbols, and actor.
The heading and prose introduce the \texttt{draft-brief} stage.
The fenced \texttt{actor} block turns that stage into an executable actor boundary.
Only the protocol header and semantic blocks are treated as executable syntax.
Prose supplies intent, rationale, and prompt context but is not interpreted as hidden control logic.

\par\noindent\begin{minipage}{\linewidth}
\begin{lstlisting}[style=plain]
---
# protocol header
protocol: research-brief-demo
version: 0.1
entry: draft-brief
symbols:
  research_question:
    type: string
  final_brief:
    type: string
    commit_policy: requires_human
actors:
  writer:
    kind: llm
---

## Draft Brief {#draft-brief}

Write a brief from curated evidence.

```actor
uses: writer
reads: [research_question]
writes: [final_brief]
```
\end{lstlisting}
\end{minipage}\par

\subsection{Protocol Header}

The protocol header defines the document-level objects that remain visible across all stages.
It gives the protocol a stable identity, identifies the entry point for execution, and declares the shared state that later blocks may read, write, or route on.
It also names reusable actors, policies, and subflows when the workflow needs them.
Some fields behave like explicit imports into the protocol namespace.
For example, \texttt{subflows} binds a local name to a child protocol package, and actor identifiers bind stage blocks to sidecar actor files.
Table~\ref{tab:xpf-header-fields} summarizes these common header fields and the role each field plays in the surrounding document.

\begin{table}[H]
\centering
\small
\renewcommand{\arraystretch}{1.12}
\begin{tabularx}{0.92\linewidth}{>{\ttfamily\raggedright\arraybackslash}p{0.22\linewidth} p{0.16\linewidth} X}
\toprule
\textnormal{\textbf{Field}} & \textbf{Required} & \textbf{Meaning} \\
\midrule
protocol & yes & Unique protocol identifier. \\
version & no & Protocol version used for compatibility and audit records. \\
entry & yes & Stage where execution begins. \\
inputs & no & External values required before the protocol can run. \\
outputs & no & Values exposed as protocol results. \\
symbols & yes & Typed workflow state declarations. \\
actors & no & Named actor profiles or sidecar actors available to stages. \\
policies & no & Reusable source, commit, or approval policies. \\
subflows & no & Child protocol packages imported under local names. \\
\bottomrule
\end{tabularx}
\caption{Common XPF protocol header fields.}
\label{tab:xpf-header-fields}
\end{table}

\subsection{Sidecar Definition Files}

The \texttt{actors} field in an XPF protocol names the actors that stages may invoke.
In the implementation, reusable actors can also be written as sidecar definition files under \texttt{agents/}.
Each file has YAML frontmatter for machine-readable invocation metadata and a Markdown body for actor-facing instructions.
Listing~\ref{lst:actor-definition-file} shows a compact example adapted from the research brief demo's \texttt{writer.agent.md} file.

\par\noindent\begin{minipage}{\linewidth}
\begin{lstlisting}[style=plain,caption={Example actor definition file.},label={lst:actor-definition-file}]
---
id: writer
kind: llm
profile: drafting-base
allowed_tools: [compose_brief]
inputs:
  research_question: str
  audience: str
  brief_goal: str
  evidence_notes: str
  final_brief: str
outputs:
  final_brief: str
---

# Actor: Writer

When `final_brief` is missing, call `compose_brief`.
Once the brief artifact exists, return `idle` and let the judge stage run.
Do not try to satisfy later revision-only checklist requirements in this stage.
\end{lstlisting}
\end{minipage}\par

The frontmatter declares the stable actor identifier, the profile used to invoke it, the tools it may call, and the symbol-shaped inputs and outputs it understands.
The Markdown body remains natural-language guidance for the actor implementation.
A stage can then bind to this file through an actor block such as \texttt{uses: writer}.
The protocol still controls which subset of the declared inputs is visible and which outputs may update workflow state.

Other sidecar directories follow the same explicit-reference pattern.
Profiles under \texttt{profiles/} provide invocation defaults such as model settings and allowed tools.
Tool files under \texttt{tools/} declare callable tool contracts, implementation references, write scopes, and artifact outputs.
Schema files under \texttt{schemas/} provide reusable structured types.
Deterministic helper code, such as a derivation rule in \texttt{rules/readiness.py}, can be referenced from a semantic block by a package-local function path.
These files act like imports, but the reference remains visible in the XPF document rather than being hidden inside prose.

\subsection{Symbol and Policy Declarations}

Symbols are declared in the header because they are protocol-level state.
A symbol declaration gives the compiler enough information to check references, validate actor outputs, and attach lifecycle behavior at runtime.

\par\noindent\begin{minipage}{\linewidth}
\begin{lstlisting}[style=plain,caption={Symbol declarations with source and commit policies.},label={lst:xpf-symbol-decls}]
symbols:
  evidence_notes:
    type: array[string]
    role: working
    source_policy: tool_or_human
    commit_policy: auto
  final_brief:
    type: string
    role: output
    source_policy: writer_only
    commit_policy: requires_human
  draft_ready:
    type: bool
    role: derived
    from: [final_brief]
\end{lstlisting}
\end{minipage}\par

The required field is \texttt{type}.
Other fields refine lifecycle behavior.
\texttt{role} distinguishes inputs, working values, derived values, judge outputs, and final outputs.
\texttt{source\_policy} constrains which producer classes may propose a value.
\texttt{commit\_policy} controls when a validated value becomes committed state.
\texttt{from} records dependencies for derived or freshness-tracked symbols.

\subsection{Stage Structure}

A Markdown heading introduces a stage.
The heading text provides the default stage name.
An explicit identifier may be used when stable references are needed.
Stage prose supplies human-readable intent and actor context.
Semantic blocks inside the stage define the executable boundary.

\par\noindent\begin{minipage}{\linewidth}
\begin{lstlisting}[style=plain,caption={Stage with actor, derivation, and flow blocks.},label={lst:xpf-stage-shape}]
## Review Brief {#review-brief}

Review the draft for correctness, coverage, and audience fit.

```actor
uses: reviewer
reads: [research_question, evidence_notes, final_brief]
writes: [review_passed, review_reason]
```

```flow
goto revise-draft when review_passed != true
goto package-brief when review_passed == true
```
\end{lstlisting}
\end{minipage}\par

A stage may contain multiple semantic blocks, but all block references are resolved against the protocol-level declarations and the stage-local boundary.

\subsection{Semantic Blocks}

XPF uses fenced blocks for declarations that affect execution.
The block type identifies how the compiler should interpret the content.

\paragraph{\texttt{actor}.} An actor block binds a stage to an actor profile and declares the symbols that may cross the actor boundary.

\par\noindent\begin{minipage}{\linewidth}
\begin{lstlisting}[style=plain]
```actor
uses: writer
reads: [research_question, evidence_notes]
writes: [final_brief]
profile: concise_writer
```
\end{lstlisting}
\end{minipage}\par

\paragraph{\texttt{derive}.} A derive block declares a deterministic or mechanical value computed from dependencies.
Simple rules can be written inline with \texttt{expr}.
More substantial rules can be imported from package-local code with \texttt{python}, using the form \texttt{module.path:function\_name}.

\par\noindent\begin{minipage}{\linewidth}
\begin{lstlisting}[style=plain]
```derive
writes: draft_ready
from: [final_brief]
expr: final_brief != null
```

```derive
writes: draft_ready
from: [final_brief]
python: rules.readiness:draft_ready
```
\end{lstlisting}
\end{minipage}\par

\paragraph{\texttt{judge}.} A judge block represents a semantic decision boundary whose outputs are still mapped into declared symbols.

\par\noindent\begin{minipage}{\linewidth}
\begin{lstlisting}[style=plain]
```judge
id: review_brief
uses: reviewer
reads: [final_brief, evidence_notes]
when: draft_ready == true
outputs:
  passed: review_passed
  reason: review_reason
```
\end{lstlisting}
\end{minipage}\par

\paragraph{\texttt{flow}.} A flow block defines control transfers. Targets must resolve to known stages or declared child protocols.

\par\noindent\begin{minipage}{\linewidth}
\begin{lstlisting}[style=plain]
```flow
goto revise-draft when review_passed != true
goto package-brief when review_passed == true
call child/protocol when needs_subtask == true
  return:
    child_output: parent_symbol
```
\end{lstlisting}
\end{minipage}\par

\paragraph{\texttt{handoff}.} A handoff block marks an explicit change in actor responsibility.

\par\noindent\begin{minipage}{\linewidth}
\begin{lstlisting}[style=plain]
```handoff
when: requires_domain_review == true
to: human_reviewer
reason: domain-specific approval required
```
\end{lstlisting}
\end{minipage}\par

\subsection{Predicates and Expressions}

Predicates in \texttt{when}, \texttt{expr}, and flow guards are intentionally small. They may reference declared symbols, constants, comparison operators, boolean connectives, and a limited set of runtime predicates such as freshness or existence checks.

\par\noindent\begin{minipage}{\linewidth}
\begin{lstlisting}[style=plain,caption={Representative predicate forms.},label={lst:xpf-predicates}]
draft_ready == true
review_passed != true
quality_score >= 0.8
exists(final_brief)
fresh(evidence_notes) and review_passed == true
\end{lstlisting}
\end{minipage}\par

The predicate language is not a general-purpose programming language. Its purpose is to make routing and readiness conditions explicit enough for parsing, dependency recovery, and runtime evaluation.

\subsection{Calls, Handoffs, and Scoped Outputs}

XPF distinguishes ordinary stage transitions from scoped child protocol calls.
A \texttt{goto} moves control to another stage in the same protocol.
A \texttt{call} enters a child protocol named in the header's \texttt{subflows} map.
The child receives its own local namespace and return boundary.
A \texttt{handoff} changes actor responsibility without hiding the change inside prompt text.

\par\noindent\begin{minipage}{\linewidth}
\begin{lstlisting}[style=plain,caption={Scoped child protocol call.},label={lst:xpf-call}]
```flow
call evidence/collect when needs_evidence == true
  with:
    topic: research_question
  return:
    collected_notes: evidence_notes
goto draft-brief when fresh(evidence_notes)
```
\end{lstlisting}
\end{minipage}\par

The target \texttt{evidence/collect} is resolved through the parent package's subflow imports rather than by scanning arbitrary prose.
Explicit return mappings prevent child protocol outputs from leaking into parent state by name coincidence.
The parent protocol decides which returned values become visible and which symbols receive them.

\subsection{Well-Formedness Rules}

An XPF document is well formed only if its structured declarations are internally consistent. The compiler enforces at least the following rules:

\begin{itemize}
  \item Every referenced symbol, stage, actor, policy, and child protocol must be declared.
  \item Actor \texttt{reads} and \texttt{writes} must be compatible with stage boundaries and symbol source policies.
  \item Actor, judge, and call outputs must map to declared symbols with compatible schemas.
  \item Derivation dependencies must resolve to declared symbols and must not introduce invalid dependency cycles.
  \item Flow targets must resolve to known stages or child protocols, and the entry stage must be reachable.
  \item Commit-gated outputs must have an explicit commit policy and a runtime path for approval or rejection.
\end{itemize}

These rules are syntax-adjacent rather than semantic. They do not prove that an actor's answer is true, complete, or persuasive. They ensure that any answer enters the workflow through declared channels.

\subsection{Complete Protocol Slice}

The following compact slice combines the main syntax elements in one stage.
It is not a full application.
It shows how symbols, actor permissions, derived readiness, review outputs, and flow predicates connect in a single readable artifact.

\par\noindent\begin{minipage}{\linewidth}
\begin{lstlisting}[style=plain,caption={Representative XPF slice from the research brief workflow.},label={lst:xpf-slice}]
symbols:
  final_brief:
    type: string
  draft_ready:
    type: bool
    role: derived
  review_passed:
    type: bool
    role: judge_output
  review_reason:
    type: string
    role: judge_output
  delivery_manifest:
    type: string
    commit_policy: requires_human

## Draft Brief

Write a brief from curated evidence. Review is a judge point.

```actor
uses: writer
reads: [research_question, audience, brief_goal, evidence_notes]
writes: [final_brief]
```

```derive
writes: draft_ready
from: [final_brief]
expr: final_brief != null
```

```judge
id: review_brief
uses: reviewer
reads: [research_question, audience, brief_goal, evidence_notes, final_brief]
when: draft_ready == true
outputs:
  passed: review_passed
  reason: review_reason
```

```flow
call revise-draft when draft_ready == true and review_passed != true
  return:
    final_brief: final_brief
goto package-brief when review_passed == true
```
\end{lstlisting}
\end{minipage}\par

\section{Compilation Phase: From Protocol Text to Typed IR}
\label{app:ir-schema}

After the protocol is authored, XFlow lowers it into a typed intermediate representation.
This phase is where the literate document becomes a runtime contract.
The compiler does not execute actors and does not judge the semantic quality of their future outputs.
Its job is narrower and more mechanical: recover the structure of the protocol, resolve names, attach policies, and reject workflows whose declared structure is inconsistent.
This makes compilation the first enforcement point for the formal/informal boundary.
Protocol prose and semantic blocks become objects that can be checked before any LLM, human actor, or tool is allowed to change workflow state.

\subsection{Parsing the Literate Surface}

Compilation begins by parsing the XPF document into three kinds of material.
The YAML frontmatter becomes the source of protocol-level declarations, including inputs, outputs, symbols, subflows, and the entry stage.
Markdown headings become stage boundaries, preserving the hierarchy that authors use to organize the workflow.
Fenced semantic blocks become typed declarations attached to the nearest stage.

This parsing step is intentionally conservative.
Ordinary prose remains documentation and actor context, but the compiler does not interpret it as a hidden command language.
This avoids a common failure mode in prompt-based systems: critical routing rules are not buried inside paragraphs that only an LLM interprets.
If a rule should affect execution, it must appear in a semantic block.

\subsection{Resolving Names and Interfaces}

Once the document is parsed, the compiler resolves names.
It checks symbol references in actor blocks, derivation rules, judge outputs, flow bindings, and handoff declarations against the symbol table.
It checks stage references in \texttt{call} and \texttt{goto} rules against the recovered stage hierarchy.
It also checks actor references against available actor specifications.
The result is a protocol graph whose edges point to known objects rather than free text labels.

This step is also where the compiler begins to enforce interfaces.
If a stage declares that an actor writes \texttt{final\_brief}, the compiler can later compare that write set with symbol policy and stage expectations.
If a judge emits \texttt{passed} into \texttt{review\_passed}, the compiler records the binding between the local output name and the global symbol.
This prevents the runtime from guessing which field in a response should be treated as the review decision.

\begin{figure}[H]
\centering
\resizebox{0.95\linewidth}{!}{%
\begin{tikzpicture}[
    node distance=0.82cm and 0.86cm,
    inputbox/.style={rectangle, rounded corners=8pt, draw=black!35, fill=black!3, minimum width=2.65cm, minimum height=0.78cm, align=center, font=\small},
    passbox/.style={rectangle, rounded corners=8pt, draw=themecolor@3, fill=themecolor@5, minimum width=2.34cm, minimum height=0.78cm, align=center, font=\small},
    gatebox/.style={rectangle, rounded corners=8pt, draw=orange!70!black, fill=orange!7, minimum width=2.38cm, minimum height=0.78cm, align=center, font=\small},
    outbox/.style={rectangle, rounded corners=8pt, draw=green!45!black, fill=green!5, minimum width=2.65cm, minimum height=0.78cm, align=center, font=\small},
    failbox/.style={rectangle, rounded corners=7pt, draw=red!55!black, fill=red!5, minimum width=3.1cm, minimum height=0.58cm, align=center, font=\scriptsize},
    arrow/.style={-{Stealth[length=2mm]}, line width=0.7pt, themecolor@3},
    reject/.style={-{Stealth[length=1.8mm]}, line width=0.6pt, dashed, red!60!black},
    stageNo/.style={circle, draw=themecolor@3, fill=white, inner sep=1.1pt, font=\scriptsize\bfseries, text=themecolor@3},
    groupLabel/.style={font=\scriptsize\bfseries, text=black!62}
]
    \node[inputbox] (xpf) {XPF source\\\scriptsize header + stages + blocks};
    \node[passbox, right=of xpf] (parse) {Parse\\\scriptsize recover surface};
    \node[passbox, right=of parse] (resolve) {Resolve\\\scriptsize names + paths};
    \node[gatebox, right=of resolve] (check) {Check\\\scriptsize interfaces + policies};
    \node[passbox, right=of check] (lower) {Lower\\\scriptsize typed objects};
    \node[outbox, right=of lower] (ir) {Typed IR\\\scriptsize runtime contract};
    \node[failbox, below=0.72cm of check] (reject) {reject malformed protocol before execution};

    \foreach \i/\n in {1/xpf,2/parse,3/resolve,4/check,5/lower,6/ir}{
        \node[stageNo, above=0.06cm of \n.north west, xshift=0.12cm] {\i};
    }
    \node[groupLabel, above=0.20cm of parse] {compiler passes};
    \node[groupLabel, above=0.20cm of xpf] {authoring artifact};
    \node[groupLabel, above=0.20cm of ir] {execution artifact};

    \draw[arrow] (xpf) -- (parse);
    \draw[arrow] (parse) -- (resolve);
    \draw[arrow] (resolve) -- (check);
    \draw[arrow] (check) -- (lower);
    \draw[arrow] (lower) -- (ir);
    \draw[reject] (check) -- node[right, font=\scriptsize] {fail fast} (reject);
\end{tikzpicture}%
}
\caption{Compilation lowers an XPF document into a typed IR package through surface parsing, name resolution, structural checking, and object lowering.}
\label{fig:appendix-compile-path}
\end{figure}
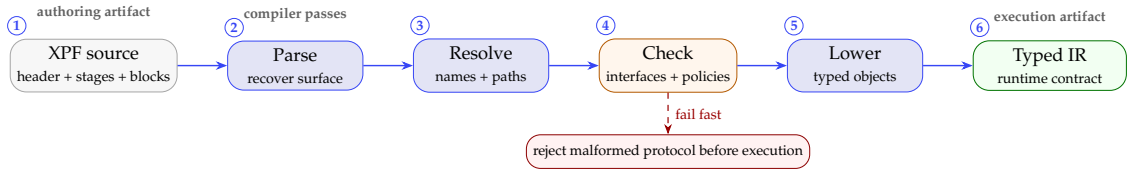

\subsection{Lowering Constructs into Runtime Objects}

The IR is not a textual mirror of the XPF file.
It is a typed object model designed for execution.
A symbol declaration becomes an \texttt{IRSymbol} with a type annotation, role, source policy, and commit policy.
A stage becomes an \texttt{IRStage} with an identifier, parent relationship, actor binding, derivations, judges, handoffs, and flow rules.
A flow rule becomes an \texttt{IRFlowRule} with a target, predicate, priority, and binding map.

The important point is that lowering changes the kind of object the system handles.
In the XPF file, \texttt{review\_passed} appears as readable text inside a block.
In the IR, it becomes a resolved symbol identifier with a declared role and validation path.
In the XPF file, \texttt{research-brief-demo/package-brief} appears as a stage path.
In the IR, it becomes a checked transition target.
This is the moment when the authoring artifact becomes an executable contract.

\par\noindent\begin{minipage}{\linewidth}
\begin{lstlisting}[style=json,caption={Conceptual IR slice for a review guarded drafting stage.},label={lst:ir-slice}]
{
  "stage_id": "research-brief-demo/draft-brief",
  "title": "Draft Brief",
  "agent": {
    "agent_ref": "writer",
    "reads": ["research_question", "audience", "brief_goal", "evidence_notes"],
    "writes": ["final_brief"]
  },
  "derives": [
    {
      "writes": "draft_ready",
      "depends_on": ["final_brief"],
      "expr": "final_brief != null"
    }
  ],
  "judges": [
    {
      "subagent_id": "review_brief",
      "agent_ref": "reviewer",
      "when": "draft_ready == true",
      "bind_outputs": {
        "passed": "review_passed",
        "reason": "review_reason"
      }
    }
  ],
  "flow_rules": [
    {"action": "call", "target": "research-brief-demo/draft-brief/revise-draft"},
    {"action": "goto", "target": "research-brief-demo/package-brief"}
  ]
}
\end{lstlisting}
\end{minipage}\par

The slice shows why the IR is useful.
The actor's read and write sets are no longer embedded in prose.
The derivation has an explicit dependency list.
The judge output names are mapped to declared symbols.
The flow rules carry targets that can be checked before execution.
None of these checks says whether the final brief is good.
They say whether the workflow has a coherent structure for producing and reviewing it.

\subsection{Static Checks as Design Commitments}

The compiler's checks follow the same design philosophy as the rest of XFlow.
They are intentionally structural.
A name resolution error means the protocol refers to something that does not exist.
A capability or policy error means an actor is being granted authority it should not have.
A flow error means control may jump to a missing or malformed stage.
A schema error means a value cannot be safely placed into its declared symbol.

These checks matter because they convert a class of runtime surprises into authoring time feedback.
If a reviewer output is bound to a missing symbol, the protocol fails early.
If a handoff exposes an undeclared field, the protocol fails early.
If a derived value depends on an unknown upstream symbol, the protocol fails early.
The runtime can then focus on the uncertain part that remains: what actors actually produce.

\subsection{A Compact Check Summary}

The table below is retained as a compact summary rather than as the main explanation.
It should be read after the preceding narrative.
Each row is a reminder of how an authoring construct becomes a checked runtime object.

\begin{center}
\small
\renewcommand{\arraystretch}{1.15}
\begin{tabularx}{0.92\linewidth}{>{\raggedright\arraybackslash}p{0.24\linewidth} >{\raggedright\arraybackslash}p{0.25\linewidth} X}
\toprule
\textbf{Authoring construct} & \textbf{IR object} & \textbf{Representative check} \\
\midrule
Symbol declaration & \texttt{IRSymbol} & Schema, role, source policy, and commit policy are attached to a resolved symbol. \\
Stage heading & \texttt{IRStage} & Stage identifiers and parent relationships are recovered from the Markdown hierarchy. \\
Actor block & \texttt{IRStageAgent} & Reads and writes refer to declared symbols and declared actor specifications. \\
Judge block & \texttt{IRSubagent} & Local judge outputs are mapped to declared workflow symbols. \\
Flow block & \texttt{IRFlowRule} & Transition targets and return bindings refer to known stages and symbols. \\
\bottomrule
\end{tabularx}
\end{center}

\section{Boundary Phase: Actor Output as a Controlled Crossing}
\label{app:actor-output}

Once a compiled protocol starts running, the most delicate moment is the crossing from actor behavior into workflow state.
Actors are intentionally flexible.
A writer may synthesize evidence, a reviewer may apply judgment, and a human operator may approve or reject a proposal.
XFlow does not attempt to formalize the internal reasoning behind those actions.
It instead controls the interface through which an actor turn becomes a runtime event.
Actors may produce open-ended content, but only structured terminal events and declared symbol writes can affect workflow state.
Everything else remains evidence, trace, or diagnostic context.

\subsection{One Actor Turn, Two Kinds of Material}

An actor turn contains both informal material and executable intent.
The informal material includes the raw response, explanatory observations, and trace summaries.
These fields are valuable because they let a reader reconstruct what happened and why.
They are not sufficient to update the protocol state.
The executable intent appears in the terminal event, whose type and payload are interpreted by the runtime.

This separation prevents the runtime from treating arbitrary prose as state.
A reviewer can write a long explanation, but the runtime looks for the declared \texttt{return} payload before updating \texttt{review\_passed} and \texttt{review\_reason}.
A publisher can describe a handoff in natural language, but the runtime activates a handoff only when a \texttt{finish} event matches a declared handoff block.
In this sense, the actor envelope is the execution-time counterpart of XPF's semantic blocks.

\par\noindent\begin{minipage}{\linewidth}
\begin{lstlisting}[style=json,caption={Compact actor output envelope.},label={lst:actor-envelope}]
{
  "trace": [
    {
      "kind": "review",
      "summary": "The draft covers the requested audience and goal.",
      "refs": [{"kind": "symbol", "id": "final_brief"}]
    }
  ],
  "observation": {
    "coverage": "required sections present",
    "risk": "minor style edits only"
  },
  "terminal": {
    "type": "return",
    "payload": {
      "outputs": {
        "passed": true,
        "reason": "The brief satisfies the declared review criteria."
      }
    }
  },
  "raw_response": "Reviewer narrative retained for audit.",
  "logs": [],
  "error": null
}
\end{lstlisting}
\end{minipage}\par

\subsection{How the Runtime Reads the Envelope}

The runtime reads the envelope in stages.
First, it records the raw response and trace so that the actor turn remains auditable.
Second, it checks whether a terminal event is present.
Third, it dispatches the terminal event according to a small vocabulary: \texttt{tool\_call}, \texttt{return}, \texttt{finish}, or \texttt{idle}.
Fourth, if the event proposes values, those values are mapped through the declarations in the current stage.
Only after that mapping succeeds can values enter the symbol lifecycle.

The \texttt{return} event in Listing~\ref{lst:actor-envelope} illustrates the path.
The actor does not write directly to \texttt{review\_passed}.
It returns a local output named \texttt{passed}.
The judge declaration from Appendix~\ref{app:xpf-syntax} maps \texttt{passed} to \texttt{review\_passed}.
This indirection is useful because it makes the boundary explicit.
The actor speaks in the local vocabulary of the terminal event, while the protocol decides which workflow symbol receives the value.

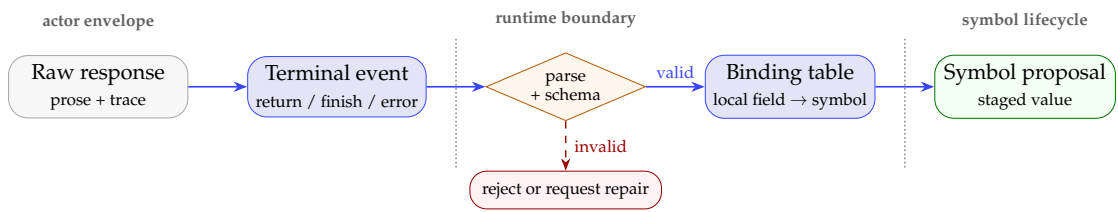
\begin{figure}[H]
\centering
\resizebox{0.94\linewidth}{!}{%
\begin{tikzpicture}[
    node distance=0.78cm and 0.86cm,
    material/.style={rectangle, rounded corners=8pt, draw=black!35, fill=black!3, minimum width=2.62cm, minimum height=0.74cm, align=center, font=\small},
    eventbox/.style={rectangle, rounded corners=8pt, draw=themecolor@3, fill=themecolor@5, minimum width=2.48cm, minimum height=0.74cm, align=center, font=\small},
    gatebox/.style={diamond, aspect=2.35, draw=orange!70!black, fill=orange!7, inner sep=1.5pt, align=center, font=\scriptsize},
    statebox/.style={rectangle, rounded corners=8pt, draw=green!45!black, fill=green!5, minimum width=2.64cm, minimum height=0.74cm, align=center, font=\small},
    failbox/.style={rectangle, rounded corners=7pt, draw=red!55!black, fill=red!5, minimum width=2.8cm, minimum height=0.55cm, align=center, font=\scriptsize},
    arrow/.style={-{Stealth[length=2mm]}, line width=0.7pt, themecolor@3},
    failarrow/.style={-{Stealth[length=1.8mm]}, line width=0.6pt, dashed, red!60!black},
    boundary/.style={densely dotted, line width=0.7pt, black!42},
    groupLabel/.style={font=\scriptsize\bfseries, text=black!62, align=center}
]
    \node[material] (raw) {Raw response\\\scriptsize prose + trace};
    \node[eventbox, right=of raw] (event) {Terminal event\\\scriptsize return / finish / error};
    \node[gatebox, right=of event] (parse) {parse\\+ schema};
    \node[eventbox, right=of parse] (bind) {Binding table\\\scriptsize local field $\to$ symbol};
    \node[statebox, right=of bind] (proposal) {Symbol proposal\\\scriptsize staged value};
    \node[failbox, below=0.72cm of parse] (blocked) {reject or request repair};

    \node[groupLabel, above=0.24cm of raw] {actor envelope};
    \node[groupLabel, above=0.24cm of parse] {runtime boundary};
    \node[groupLabel, above=0.24cm of proposal] {symbol lifecycle};
    \draw[boundary] ($(event.east)!0.5!(parse.west)+(0,0.70)$) -- ($(event.east)!0.5!(parse.west)+(0,-1.18)$);
    \draw[boundary] ($(bind.east)!0.5!(proposal.west)+(0,0.70)$) -- ($(bind.east)!0.5!(proposal.west)+(0,-1.18)$);

    \draw[arrow] (raw) -- (event);
    \draw[arrow] (event) -- (parse);
    \draw[arrow] (parse) -- node[above, font=\scriptsize] {valid} (bind);
    \draw[arrow] (bind) -- (proposal);
    \draw[failarrow] (parse) -- node[right, font=\scriptsize] {invalid} (blocked);
\end{tikzpicture}%
}
\caption{Actor output crosses into workflow state in three steps: a terminal event is parsed, its fields are bound to declared symbols, and only then is a staged symbol proposal created.}
\label{fig:actor-boundary-path}
\end{figure}

\subsection{Terminal Events as a Small Vocabulary}

The terminal vocabulary is deliberately small.
A \texttt{tool\_call} asks the runtime to execute a named tool with validated arguments.
A \texttt{return} event proposes values for stage or judge outputs.
A \texttt{finish} event submits fields for a handoff declared in the protocol.
An \texttt{idle} event records that the actor has no executable action in the current turn.

This vocabulary does not limit what actors can think or say.
It limits what the runtime has to execute.
That distinction matters for reliability.
The system can preserve a detailed raw response for audit while still refusing to infer a state transition from prose alone.
If the terminal event is malformed, missing, or inconsistent with the current stage declaration, the actor turn can be recorded without being treated as committed state.

The envelope therefore preserves rich actor context without granting that context authority.
A reviewer can inspect the raw response and trace.
The runtime relies only on terminal events and schema-checked values.

\subsection{From Boundary Crossing to Symbol Lifecycle}

After the terminal event has been parsed and bound to workflow symbols, the proposed values move into the lifecycle described in Appendix~\ref{app:session-format}.
This is where schema checks, source policies, and commit policies take effect.
The actor boundary therefore does not end with parsing JSON.
It ends only when the value has been rejected, blocked for approval, or admitted into the symbol store under a known state.

This design gives XFlow a clear answer to a common multi-agent question: when does an actor output become something other actors may rely on?
The answer is not ``when the model said it,'' and it is not ``when the text looked plausible.''
The answer is: when the actor output has crossed the declared event boundary.
It must also match a protocol binding, pass validation, and satisfy its commit policy.

\section{Execution Phase: Staging Uncertainty in Symbol State}
\label{app:session-format}

The Execution phase begins after an actor output has crossed the boundary described in Appendix~\ref{app:actor-output}.
At this point the system may have a candidate value, but it still does not necessarily have committed workflow state.
XFlow treats this gap as a central design problem.
Values move through a lifecycle that records uncertainty, validation, approval, freshness, and commit status.
No actor output becomes shared truth simply because an actor produced it.

\subsection{A Running Example: Packaging a Brief}

Consider the \texttt{delivery\_manifest} symbol from the research brief workflow.
The delivery packager may produce a manifest after the brief has passed review, but the symbol carries \texttt{commit\_policy: requires\_human}.
This means the system can accept a structurally valid proposal without immediately treating it as durable output.
The value can be visible to an operator, inspected against artifacts, and either approved or rejected.

This example shows why the lifecycle is more expressive than a simple variable assignment.
A normal assignment would replace the old value with the new one.
XFlow instead records a proposed value, checks whether the producer is allowed to propose it, validates the shape of the value, and then pauses if the commit policy requires human approval.
The workflow can therefore distinguish between ``a value has been suggested,'' ``a value has the right structure,'' and ``a value is approved as shared truth.''

\par\noindent\begin{minipage}{\linewidth}
\begin{lstlisting}[style=json,caption={Compact symbol cell snapshot for a human gated deliverable.},label={lst:symbol-cell}]
{
  "symbol_id": "delivery_manifest",
  "value": "{... proposed package manifest ...}",
  "has_value": true,
  "role": "working",
  "state": "pending_human",
  "freshness": "fresh",
  "source_policy": "llm_proposable",
  "commit_policy": "requires_human",
  "provenance_refs": ["actor_turn:delivery_packager:7"],
  "version": 3,
  "committed_value": null,
  "committed_version": null
}
\end{lstlisting}
\end{minipage}\par

\subsection{Lifecycle as Increasing State Authority}

The core path is \textsc{Unknown} $\rightarrow$ \textsc{Proposed} $\rightarrow$ \textsc{Validated} $\rightarrow$ \textsc{Committed}.
The path should be read as increasing authority over a value.
In \textsc{Unknown}, the protocol knows that a symbol exists but has no value.
In \textsc{Proposed}, a candidate has arrived and can be attributed to a producer.
In \textsc{Validated}, the candidate has passed structural checks.
In \textsc{Committed}, the runtime has taken a durable snapshot that downstream stages and recovery logic may rely on.

The auxiliary states make the lifecycle operational.
\textsc{Pending Human} represents a valid value that is waiting for an approval decision.
\textsc{Invalid} represents a candidate that failed schema validation, policy checks, or approval.
These states are not cosmetic.
They allow the runtime and the operator interface to show why progress has stopped and what kind of intervention is required.

\begin{figure}[H]
\centering
\resizebox{0.96\linewidth}{!}{%
\begin{tikzpicture}[
    node distance=0.74cm and 0.78cm,
    state/.style={rectangle, rounded corners=8pt, draw=themecolor@3, fill=themecolor@5, minimum width=2.05cm, minimum height=0.66cm, align=center, font=\small\scshape},
    gate/.style={rectangle, rounded corners=8pt, draw=orange!70!black, fill=orange!7, minimum width=2.12cm, minimum height=0.66cm, align=center, font=\small\scshape},
    bad/.style={rectangle, rounded corners=8pt, draw=red!55!black, fill=red!5, minimum width=2.0cm, minimum height=0.60cm, align=center, font=\small\scshape},
    freshbox/.style={rectangle, rounded corners=8pt, draw=green!45!black, fill=green!5, minimum width=2.08cm, minimum height=0.62cm, align=center, font=\small},
    lane/.style={font=\scriptsize\bfseries, text=black!62, align=right},
    arrow/.style={-{Stealth[length=2mm]}, line width=0.7pt, themecolor@3},
    gatearrow/.style={-{Stealth[length=1.9mm]}, line width=0.65pt, dashed, orange!75!black},
    failarrow/.style={-{Stealth[length=1.9mm]}, line width=0.65pt, dashed, red!60!black},
    sep/.style={densely dotted, black!30}
]
    \node[lane] (authLabel) {authority\\state};
    \node[state, right=0.58cm of authLabel] (unknown) {Unknown};
    \node[state, right=of unknown] (proposed) {Proposed};
    \node[state, right=of proposed] (validated) {Validated};
    \node[state, right=2.25cm of validated] (committed) {Committed};
    \coordinate (commitMid) at ($(validated)!0.5!(committed)$);
    \node[gate, above=0.74cm of commitMid] (pending) {Pending Human};
    \node[bad, below=0.74cm of validated] (invalid) {Invalid};

    \draw[arrow] (unknown) -- (proposed);
    \draw[arrow] (proposed) -- (validated);
    \draw[arrow] (validated) -- (committed);
    \draw[gatearrow] (validated.north) |- (pending.west);
    \draw[arrow] (pending.east) -| (committed.north);
    \draw[failarrow] (proposed.south) |- (invalid.west);
    \draw[failarrow] (pending.south) |- (invalid.east);

    \draw[sep] ($(authLabel.south west)+(0,-0.55)$) -- ($(committed.south east)+(0.10,-0.55)$);

    \node[lane, below=1.65cm of authLabel] (freshLabel) {dependency\\freshness};
    \node[freshbox, right=0.58cm of freshLabel] (freshNode) {Fresh};
    \node[freshbox, right=of freshNode] (staleNode) {Stale};
    \node[freshbox, right=of staleNode] (recomputeNode) {Recomputed};
    \draw[gatearrow] (freshNode) -- (staleNode);
    \draw[arrow] (staleNode) -- (recomputeNode);
    \coordinate (freshReturn) at ($(freshNode.west)+(-0.42,0)$);
    \draw[arrow] (recomputeNode.south) -- ++(0,-0.36) -| (freshReturn) -- (freshNode.west);
\end{tikzpicture}%
}
\caption{Symbol lifecycle separates two concerns: authority state decides whether a value may be trusted as committed state, while freshness records whether derived values still match their dependencies.}
\label{fig:symbol-lifecycle-appendix}
\end{figure}
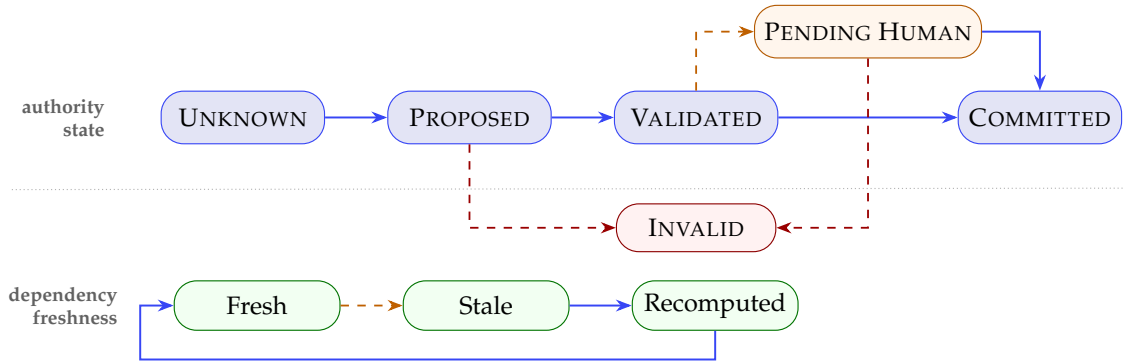

\subsection{Policies and Freshness}

Two policy dimensions govern the transition from actor output to shared state.
Source policy asks whether the producer is allowed to propose the value.
Commit policy asks what boundary must be crossed before the value becomes durable.
A value can therefore be structurally valid and still not be committed.
This is exactly what happens with \texttt{delivery\_manifest}: the value may pass schema validation, but the commit policy still requires human approval.

Freshness is tracked separately because a value can be valid but no longer current.
If \texttt{final\_brief} changes after a revision, then derived values depending on it may need to be recomputed.
XFlow marks dependent derived symbols as stale rather than silently reusing old results.
When the derivation runs again, the symbol receives a new version and becomes fresh.
This separates two questions that are often mixed together: whether a value has the right shape, and whether it is up to date with respect to its dependencies.

Lifecycle state captures authority over a value: proposed, validated, pending, committed, or invalid.
Freshness captures whether a computed value still reflects the versions of the symbols it depends on.

\subsection{Transactions and Failure Behavior}

The runtime applies updates transactionally.
A batch of proposed symbol changes is checked before downstream operations observe it.
If a required schema check fails, the batch does not partially update committed state.
If a human gate is required, the session records a blocked request rather than silently proceeding.
If an approval is denied, the proposal is marked invalid and the protocol can retry, revise, or stop according to its declared flow.

This transactional discipline is what makes staged uncertainty practical.
The system can tolerate uncertain actors because it does not confuse actor output with committed state.
It can preserve the proposed material for audit, expose it to an operator, and still protect downstream stages from relying on a value before the declared thresholds have been crossed.

\section{Recovery Phase: Session State and Queryable Records}
\label{app:recovery-reference}

The final phase concerns what happens when execution pauses, fails, or needs to be inspected.
Long-running multi-actor workflows rarely complete in a single uninterrupted process.
They may wait for a human reviewer, pause for tool approval, branch into parallel work, or resume on another machine.
XFlow therefore treats session persistence and queryable execution records as part of the trust boundary rather than as optional logging.
Across pauses and restarts, a resumed workflow reloads typed frames, symbol cells, pending human gates, actor turns, artifacts, and event history instead of relying on an informal transcript.

\subsection{What a Session Must Remember}

A session must remember more than the latest output.
It must remember where the workflow is, which frame is active, and which symbols are known.
It must also remember which values are pending approval, which handoffs have been activated, and which actor turns produced the current state.
Without this information, recovery would collapse back into prompt-based continuation.
The system would have to summarize a transcript and hope that the next actor reconstructs the right state.

XFlow avoids that failure mode by persisting structured session state.
The session manifest stores run status, frame stack, pending blocked requests, handoffs, and parallel regions.
Frame files store per-frame runtime state and symbol cells.
Actor turn logs preserve prompts, raw responses, traces, terminal events, logs, and errors.
Artifact ledgers record which files were produced, where they are stored, and which stage owns them.
The important design choice is that each record answers a different recovery question.

\par\noindent\begin{minipage}{\linewidth}
\begin{lstlisting}[style=json,caption={Compact session manifest for a blocked human approval.},label={lst:session-manifest}]
{
  "status": "blocked",
  "root_protocol_id": "research-brief-demo",
  "frame_stack": ["frame:intake", "frame:package-brief"],
  "pending_blocked": {
    "kind": "commit_approval",
    "stage_id": "package-brief",
    "actor_id": "delivery_packager",
    "commit_outputs": ["delivery_manifest"]
  },
  "activated_handoffs": [],
  "parallel_regions": {},
  "commit_required_checks": ["delivery_manifest"]
}
\end{lstlisting}
\end{minipage}\par

The manifest above says that the session is not merely incomplete; it is incomplete for a specific formal reason.
It is blocked on commit approval for \texttt{delivery\_manifest}.
That distinction matters for operators.
A blocked judge, a tool approval, a human input request, and a parallel wait all require different interventions.
By serializing the blocked request as structured state, XFlow makes the next action explicit.

\subsection{Recovery as Structured Reconstruction}

Recovery begins by loading the compiled protocol package.
This step reestablishes the structural contract: stages, symbols, policies, actor bindings, flow rules, and handoff declarations.
The runtime then restores the frame stack and current stage identifiers, which gives control a declared place to resume.
Symbol cells are restored with values, states, freshness, versions, provenance, and commit metadata.
Pending blocked requests are restored so that human gates remain visible rather than being lost in a process restart.

The actor log and event log then provide history.
They are not used as vague memory.
They are used as auditable evidence explaining how the session reached its current state.
An operator can inspect the actor turn that proposed a value.
They can also inspect the terminal event that requested a return or handoff, the symbol update that followed validation, and the commit event that made a value durable.
Recovery therefore preserves both sides of the boundary: structured runtime state and informal actor evidence.

\subsection{Operator View Over Session State}

An operator view should be understood as a projection over the same session state, not as a separate execution mechanism.
Its purpose is to make blocked requests, symbol states, actor turns, frame stack, handoffs, and events visible when an operator needs to intervene.
The important design point is that each visible item is already represented as structured runtime state.
Blocked requests identify formal gates, symbol cells carry freshness and commit status, and approval decisions are durable runtime events.

For the paper, the concrete interface is less important than the questions that the recorded state must answer.
The operator should be able to determine what the session is waiting for and which value is being proposed.
They should also see which actor produced it, which symbols are stale, which events led here, and which branch is blocked.
These questions mirror the same staged structure used throughout the system.

\begin{center}
\small
\renewcommand{\arraystretch}{1.15}
\begin{tabularx}{0.88\linewidth}{>{\ttfamily\raggedright\arraybackslash}p{0.30\linewidth} X}
\toprule
\textnormal{\textbf{Runtime record}} & \textbf{Operational question} \\
\midrule
blocked request & What formal gate is stopping progress? \\
event log & Which runtime events led to the current state? \\
symbol freshness & Which derived values need recomputation? \\
pending judgment & How can a blocked judgment be supplied? \\
commit gate & Which pending value should become committed state? \\
branch frame & Which branch-local state should be inspected? \\
\bottomrule
\end{tabularx}
\end{center}

\subsection{Execution Records as Queryable History}

Session files support recovery, while execution records support inspection across time.
XFlow can export sessions, events, actor turns, frames, symbol history, parallel branch state, violations, and sparse metrics to a queryable store.
This makes a workflow run analyzable after the fact.
Instead of asking what the transcript seems to imply, a reviewer can query the recorded state directly.
They can see which symbol changed, which actor produced the candidate, which validation event occurred, and which commit event finalized the value.

This design is especially important for parallel and human-in-the-loop workflows.
In a parallel region, an operator may need to distinguish a failed branch from a branch that was ignored after a join threshold was met.
In a human approval path, an auditor may need to distinguish a value that was proposed from a value that was committed.
The recorded events preserve these distinctions explicitly.
The result is not just logging for debugging; it is a durable record of the formal/informal boundary in action.

Taken together, the appendices follow the execution pipeline.
XPF authors the contract.
The compiler lowers it to typed IR.
The actor envelope controls boundary crossing.
The symbol lifecycle stages uncertainty.
Session recovery preserves those guarantees over time.

\end{document}